\documentclass[fleqn,usenatbib]{mnras}
\pdfoutput=1

\usepackage[T1]{fontenc}
\usepackage{ae,aecompl}

\usepackage{newtxtext,newtxmath}
\usepackage{amsmath, amssymb}
\usepackage{graphicx}
\usepackage{epsfig}
\usepackage{enumerate}

\newcommand{\be}{\begin{equation}}
\newcommand{\ee}{\end{equation}}

\newcommand      \tabby {KIC 8462852}

\begin{document}
\title[Tidally Detached Exomoons]{Orphaned Exomoons: Tidal Detachment and Evaporation Following an Exoplanet-Star Collision}

\author[Martinez, Stone, Metzger]{Miguel Martinez$^1$, Nicholas C. Stone$^{1, 2, 3}$, Brian D. Metzger$^1$\\
$^1$Department of Physics and Columbia Astrophysics Laboratory, Columbia University, New York, NY 10027, USA\\
$^2$Racah Institute of Physics, The Hebrew University, Jerusalem, 91904, Israel \\
$^3$Department of Astronomy, University of Maryland, College Park, MD 20742, USA}

\maketitle

\begin{abstract}
Gravitational perturbations on an exoplanet from a massive outer body, such as the Kozai-Lidov mechanism, can pump the exoplanet's eccentricity up to values that will destroy it via a collision or strong interaction with its parent star.  During the final stages of this process, any exomoons orbiting the exoplanet will be detached by the star's tidal force and placed into orbit around the star.  Using ensembles of three and four-body simulations, we demonstrate that while most of these detached bodies either collide with their star or are ejected from the system, a substantial fraction, $\sim 10\%$, of such "orphaned" exomoons (with initial properties similar to those of the Galilean satellites in our own solar system) will outlive their parent exoplanet.  The detached exomoons generally orbit inside the ice line, so that strong radiative heating will evaporate any volatile-rich layers, producing a strong outgassing of gas and dust, analogous to a comet's perihelion passage.  Small dust grains ejected from the exomoon may help generate an opaque cloud surrounding the orbiting body but are quickly removed by radiation blow-out.  By contrast, larger solid particles inherit the orbital properties of the parent exomoon, feeding an eccentric disk of solids that drains more gradually onto the star via Poynting-Robertson drag, and which could result in longer-timescale dimming of the star.  For characteristic exomoon evaporation times of $\sim 10^{5}-10^{6}$ yr, attenuation of the stellar light arising from one or more out-gassing exomoons provides a promising explanation for both the dipping and secular dimming behavior observed from \tabby\,(Boyajian's Star).
 \end{abstract}

\section{Introduction}
The observed diversity of exoplanetary architectures suggests that many exoplanet systems have undergone phases of complex dynamical evolution.  This is attested to by the high observed eccentricities of many exoplanet orbits (e.g.~\citealt{Rasio&Ford96,Juric&Tremaine08,Kane+12, Winn&Fabrycky15}); the presence of hot gas giant exoplanets on orbits which are misaligned$-$or even counter-rotating$-$with respect to the spin axes of their stars (\citealt{Fabrycky&Tremaine07,Naoz+11}); the existence of interstellar exoplanets (e.g.~\citealt{Schneider+16,Mroz+18}) or exoplanetesimals (e.g.~\citealt{Raymond+18,Rafikov18}) which were dynamically ejected from their birth systems; and by observations of dusty debris rings from exoplanetary collisions (e.g.~\citealt{Song+05,Melis+13,Kenyon&Bromley16}).  Although much of this behavior takes place soon after stellar birth, other dynamic processes can operate over longer timescales covering the main sequence, or even post-main sequence, lifetime.  

An outer stellar, or sub-stellar, companion on a long-period orbit will exert gravitational torques on the inner exoplanet, which can generate secular oscillations of its orbital properties (\citealt{Kozai62,Lidov62}; see \citealt{Naoz16} for a review).  In some cases the exoplanet's eccentricity will be driven to sufficiently high values to strongly interact$-$or even physically collide$-$with the central star (\citealt{Wu&Murray03,Fabrycky&Tremaine07,Stephan+18}).  For a circular outer binary, the inner exoplanet is driven to such a collision only for a narrow range of the mutual orbital inclinations. %$i \approx 39^{\circ}$ or $141^{\circ}$.  
However, in the more general case, when the outer perturber possesses non-zero eccentricity, the inner exoplanet's eccentricity can reach arbitrarily high values, even in the nearly co-planar regime (\citealt{Naoz+11,Naoz+12,Li+14,Hamers+16}).

Such eccentricity oscillations may be relevant for the {\it Kepler}-field star \tabby, which has exhibited several highly unusual dimming events of variable depth and duration \citep{Boyajian+16}, with temporary reductions in the total stellar flux ranging from $0.5-20$\%.  These dips, which continue to the present day \citep{Boyajian+18}, likely arise from transits of the star by localized, optically-thick clouds of dusty debris, as might be produced by a "giant swarm" of comets \citep{Boyajian+16,Bodman&Quillen16} or a smaller number of more massive outgassing bodies \citep{Neslusan&Budaj17}.  In addition to these short-timescale dimming events, archival observations of \tabby~over the century from 1890 to 1989 show more gradual (``secular'') fading of its flux by 14\% (\citealt{Schaefer16}; but see also \citealt{Castelaz&Barker18}), as well as by another 3\% over the 4 year {\it Kepler} mission \citep{Montet&Simon16}.  Ground-based observations, taken in the years both before and after {\it Kepler}, confirm this overall secular decay and show that there is structure in the light curve over a wide range of timescales \citep{Schaefer+18}, including brief {\it increases} in brightness \citep{Simon+18}.  

Among several potential explanations for the behavior of \tabby, \citet{Wright&Sigurdsson16} discussed a exoplanetary collision with the star.  This possibility was explored by \citet{Metzger+17}, who showed that the observed secular-timescale dimming of \tabby~could be the result of the stellar flux returning back to its original, pre-collision level following the collision, which would naturally produce an (unobserved) brightening due to the sudden injection of energy into the star's surface layers from the sinking exoplanet.  One difficulty of this model is the implied high occurrence rate of star-planet collisions needed to explain the existence of even a single \tabby-like system in the limited {\it Kepler} field of $\sim 10^{5}$ stars.  The predicted duration of the dimming phase is very short ($\sim 10-10^{3}$ yr, depending on the victim exoplanet's mass), compared to the star's main sequence lifetime $\approx 2\times 10^{9}$ yr.  A second difficulty is the subsequently observed non-monotonicity of the secular dimming \citep{Simon+18}, which is hard to reconcile with the smooth contraction of the star's outer layers to the main sequence.

Interestingly, \citet{Wyatt+18} showed that the secular dimming, much like the dips themselves, could result from obscuration by dust distributed nearly continuously along an elliptical orbit about the star, with pericentre distance $q \sim 0.03-0.6$ AU, eccentricity $e \gtrsim 0.7-0.97$, and semi-major axes $a = q/(1-e) \gtrsim 1-2$ AU.  These orbital parameters were argued to be consistent with the observed dip durations, the fraction of the star's light obscured by the debris needed to explain the secular decay,  and upper limits on reprocessed infrared emission (\citealt{Schaefer+18}; their Fig.~4).  

In addition to their stellar structure calculations, \citet{Metzger+17} described two mechanisms by which the same exoplanet-star collisions could give rise to dusty debris on eccentric orbits around the star.  Firstly, if the average density of the exoplanet is less than that of the star,
then the exoplanet will be tidally disrupted outside of the stellar surface \citep{Guillochon+11,Metzger+12} prior to collision.  This process will place a fraction of the exoplanetary debris into orbit around the star, with a wide range of semi-major axes.  However, the debris would share a common pericenter distance of at most a few stellar radii, $\sim 0.01-0.02$ AU, making it unclear whether it would maintain sufficient ``clumpiness" to explain the discrete dips seen in \tabby~ (given the strong tidal forces, and periodic passage of the material inside the sublimation radius for silicate rock of $\sim 0.1$ AU - see Eq.~\ref{eq:Rsub}).  

\citet{Metzger+17} proposed a second, potentially more promising, origin for dusty debris from a exoplanet-star collision: a tidally-detached exomoon.  As the pericenter of the exoplanet decreases, its pericentric Hill sphere\footnote{By this, we mean the size of the Hill sphere at the exoplanet's orbital pericenter, where it is smallest.} also shrinks until eventually any exomoon in orbit around it will feel the gravitational force of the star, which can unbind the exomoon and place it onto a new orbit about the star.  If the exomoon survives the detachment process and subsequent gravitational interactions with its parent exoplanet, then the exomoon could remain in a stable orbit around the star.  

The pericenters of such ``orphaned'' exomoons are sufficiently small that they will experience strong irradiation from the star, sublimating any volatile-rich surface layers, and resulting in massive outgassing of gas and dust.  This could plausibly generate an opaque cloud of debris surrounding the exomoon, providing a screen for generating short-lived dips in the observed light curve of the star as the exomoon's orbit passed in front of the observer line-of-sight  (in analogy to e.g. \citealt{Rappaport+12, Rappaport+14, SanchisOjeda+15}).  
Solid particles released from exomoon outgassing would feed a ring of debris extending over most of the exomoon's orbit, which could give rise to longer,  secular-timescale dimming (or brightening) of the star, similar to the scenario envisioned by \citet{Wyatt+18}.

In this paper we explore the tidally-detached exomoon scenario in greater detail.  In $\S\ref{sec:dynamics}$ we examine the dynamical excitation of an exoplanet onto the type of high-eccentricity orbit that can collide with the central star.  While many of the subsequent arguments in this paper could apply to alternative high-$e$ migration channels, for the sake of concreteness, we focus primarily on the Kozai-Lidov effect.  After identifying star-crossing orbits, we initialize ``zoomed-in" simulations on the final stages of the collision, with the exomoon initially in orbit around the exoplanet, to explore its range of possible fates.  We focus on the properties of the surviving exomoon orbits resulting from tidal detachment and compare them to those permitted by observations of \tabby.  In $\S\ref{sec:outgassing}$ we examine the photo-evaporation of volatile-rich exomoons and the evolution of their solid debris under the influence of Poynting-Robertson drag from the central star.  In \S\ref{sec:discussion} we discuss the implications of our results on outgassing rates and observable lifetimes for the debris from tidally detached exomoons, focusing particularly on the orbital and exomoon parameters needed to explain the observed secular dimming in \tabby.  We assess the fraction of all stars which must experience a close exoplanetary encounter for our mechanism to represent a viable explanation for \tabby.  In \S\ref{sec:conclusions} we summarize our conclusions.  Table \ref{table:variables} provides a list of variables commonly used throughout the text.

\begin{table*}
\caption{Commonly Used Variables \label{table:variables}}
%\centering
\begin{tabular}{ccccccccccc}
\hline\hline
Variable & Definition\\
\hline
$\{M_{\star},R_{\star},L_{\star},T_{\star}, \mathcal{T}_{\star}\}$ & Mass, Radius, Luminosity, Temperature, and Rotational Period of Central Star\\
$\{M_{\rm o}, a_{\rm o}, e_{\rm o}\}$ & Mass, Semimajor Axis, and Eccentricity of Outer Perturber\\
$\{M_{\rm p}, R_{\rm p}\}$ & Mass and Radius of Exoplanet\\
$\{M_{\rm m}, R_{\rm m}\}$ & Mass and Radius of Exomoon\\
$\{a_{\rm p}, e_{\rm p}, i_{\rm p}, \Omega_{\rm p}, \omega_{\rm p}, f_{\rm p} \}$ & Heliocentric Orbital Elements of Exomoon-Bearing Exoplanet \\
$\{a_{\rm m}, e_{\rm m}, i_{\rm m}, \Omega_{\rm m}, \omega_{\rm m}, f_{\rm m} \}$ & Pre-detachment Planetocentric Orbital Elements of Exomoon \\
$\{\tilde{a}_{\rm m}, \tilde{e}_{\rm m}, \tilde{i}_{\rm m}, \tilde{\Omega}_{\rm m}, \tilde{\omega}_{\rm m}, \tilde{f}_{\rm } \}$ & Post-detachment Heliocentric Orbital Elements of Exomoon \\
$r_{\rm H}$ & Hill Radius of the Exoplanet \\
$\{\tau_{\rm KL}^{\rm quad},\tau_{\rm KL}^{\rm oct}\}$ & Quadropole- and Octopole-Order Kozai-Lidov Timescales \\
$\langle t_{\rm sub} \rangle$ & Orbit-Averaged Sublimation Time of the Detached Exomoon \\
$\{t_{\rm PR}^{\rm circ},t_{\rm PR}^{\rm rad}\}$ & Poynting-Robertson Drag Infall Timescales for Dust in Circular and Radial Orbits \\
$b$ & Dust Grain Radius \\
$b_{\rm rad}$ & Dust Grain Size Lower Limit Due to Radiation Blowout\\
$b_{\rm ss}$ & Maximum Size of Dust Grains in Steady State PR Inflow \\
$t_{\rm solid}$ & Lifetime of Dust Grains in the Solid Disk Set by PR Drag \\
$t_{\rm age}$ & Actual Age of the Solid Disk \\
\hline\\
\end{tabular}
\end{table*}

\section{Dynamics of Exomoon Detachment}
\label{sec:dynamics}
We consider an initially hierarchical system of four bodies: the primary star, of mass $M_\star$ and radius $R_\star$; the distant secondary perturber, with mass $M_{\rm o}$ on an orbit with semi-major axis $a_{\rm o}$ and eccentricity $e_{\rm o}$; the exoplanet subject to secular perturbations, with mass $M_{\rm p}$ and radius $R_{\rm p}$; and an exomoon orbiting the exoplanet, with mass $M_{\rm m}$ and radius $R_{\rm m}$.  The orbital elements of the exoplanet around the primary are $\{a_{\rm p}, e_{\rm p}, i_{\rm p}, \Omega_{\rm p}, \omega_{\rm p}, f_{\rm p} \}$, while those of the exomoon around the exoplanet are $\{a_{\rm m}, e_{\rm m}, i_{\rm m}, \Omega_{\rm m}, \omega_{\rm m}, f_{\rm m} \}$.  We assume that $M_\star \gg M_{\rm p} \gg M_{\rm m}$, and that each of the nested hierarchical orbits is (at least initially) Hill-stable, and far from mean motion resonance.  In addition, all orbital inclinations are defined with respect to the primary-secondary orbital plane.

For the remainder of the paper, we will keep analytic formulae as general as possible, but whenever we present results, we model the primary star using the best-fit parameters for \tabby{}~from \citet{Boyajian+16}: mass $M \approx 1.43M_{\odot}$, radius $R_{\star} \approx 1.58 R_{\odot}$, luminosity $L_{\star} \approx 4.7L_{\odot}$, and rotational period $\mathcal{T}_\star \approx 0.88~{\rm d}$ (though see \citealt{Makarov&Goldin16}).

\subsection{Kozai-Lidov Oscillations}
\label{sec:Kozai}

While massive exoplanets are thought to form in quasi-circular orbits, a number of dynamical mechanisms exist to excite their eccentricities to values $e_{\rm p}\approx 1$.  In this section, we focus on the Kozai-Lidov mechanism (\citealt{Lidov62}; \citealt{Kozai62}), which is a promising way to drive an exoplanet to high eccentricity over timescales comparable to the stellar main-sequence lifetime.  Qualitatively different processes, such as strong planet-planet scatterings \citep{Rasio&Ford96} and secular chaos \citep{Wu&Lithwick11, Hamers+17}, may produce the same end result, but we do not investigate these in detail here.

In our scenario, the outer massive perturber is a stellar or sub-stellar companion.  For instance, \citet{Boyajian+16} detected an M-dwarf companion star (of assumed mass $M_{\rm o} \simeq 0.4M_{\odot}$) at an angular distance of 1.96 arcsec from \tabby, corresponding to a physical distance of $\approx 900$ AU if the M-dwarf resides at the same distance.  Although proper motion observations of the companion indicate that it is actually unbound from \tabby~\citep{Clemens+18}, and its positional coincidence is a chance superposition, this conclusion is called into question by recent {\it Gaia} position measurements (D.~Clemens, private communication).  
%Although the status of the observed M-dwarf appears to be an open question at the time of writing, unobserved sub-stellar mass objects such as brown dwarfs or gas giants should also be capable of inducing Kozai-Lidov cycles.

Under the standard quadrupole-order expansion, appropriate for circular outer binary orbits, the characteristic Kozai-Lidov timescale (e.g.~\citealt{Liu+15}, their Eq.~21) is given by
\begin{eqnarray}
&& \tau_{\rm KL}^{\rm quad} \approx 5.5\,\,{\rm Myr}\left(\frac{a_{\rm p}}{20\,{\rm AU}}\right)^{-3/2} \times \nonumber \\
&& \left(\frac{M_{\rm o}}{0.4M_{\odot}}\right)^{-1}\left(\frac{M_{\star}}{1.43M_{\odot}}\right)^{1/2}\left(\frac{a_{\rm o}}{10^{3}{\rm AU}}\right)^{3}(1-e_{\rm o}^{2})^{3/2},
\end{eqnarray}
where $M_{\star} \simeq 1.43M_{\odot}$ is normalized to the mass of \tabby.  This timescale is much shorter than the age of \tabby, which shows that the normal circular, quadrupole-order Kozai-Lidov mechanism would result in the exoplanetary collision happening much earlier in its evolution than its present age.  Furthermore, the initial orbital misalignment between the exoplanet and the perturber would need to be fine-tuned in order to excite the exoplanet into a direct stellar collision.

\begin{figure*}
\includegraphics[width=1.0\textwidth]{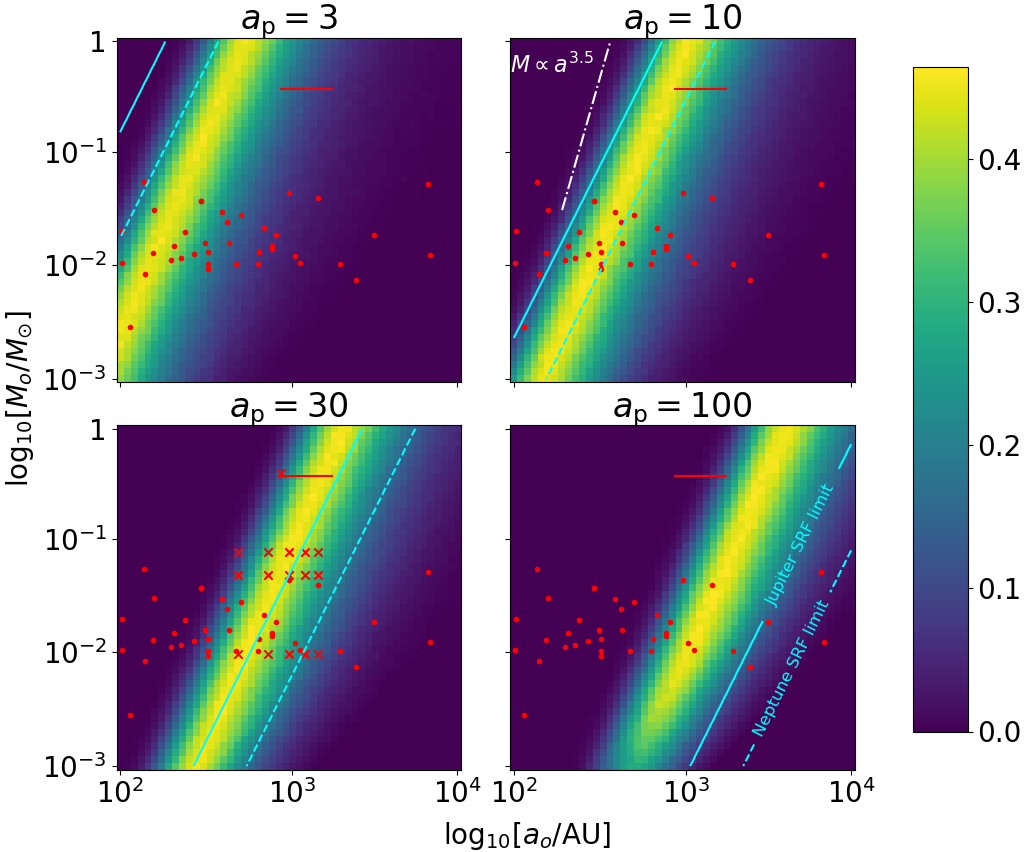}

\caption{A range of outer perturber properties can give rise to a exoplanet-star collision via the Kozai-Lidov mechanism over the main sequence lifetime of stars like \tabby.  The color scale shows the fraction of systems with exoplanetary orbit flip timescales in the range $2\times 10^{8}-2\times 10^{9}$ yrs, as a function of the semi-major axis $a_{\rm o}$ and mass $M_{\rm o}$ of the outer perturber.  These are calculated by means of a Monte Carlo method in which the outer eccentricity, mutual inclination, and mutual longitude of pericenter are sampled from thermal, isotropic, and uniform distributions, respectively.  Different panels show the results for different assumed values of the semi-major axis of the exoplanet (in all cases we take the initial exoplanet eccentricity $e_{\rm p} = 0.05$).  A horizontal solid red line shows the estimated properties of the possible M-dwarf companion of \tabby~\citep{Boyajian+16}, while red dots show the properties of directly imaged exoplanet and brown companions for other stars.  Red crosses indicate the perturber properties used in our fiducial model simulations in $\S\ref{sec:detachment}$.  Blue lines indicate the limiting $M_{\rm o}(a_{\rm o})$ for which short-range forces (primarily tides) arrest Kozai cycles prior to a planet-star collision (arrest may occur in the parameter space underneath these curves, which take the median of the thermal distribution, $e_{\rm o}=2^{-1/2}$); solid blue represents a Jovian victim planet, and dashed blue a Neptunian one.  A white line in the upper right panel shows, for illustration, the power-law dependence $M_{\rm o} \propto a_{\rm o}^{7/2}$ predicted from the analytic octopole-order KL timescale (Eq. \ref{eq:tauKL}).}
\label{fig:KLtime}
\end{figure*}

More promising is the eccentric Kozai-Lidov mechanism, which occurs when octupole contributions to the binary potential become important -  for example, when the outer binary has finite eccentricity ($e_{\rm o} \ne 0$).  This operates over a longer timescale and can reduce the exoplanet pericenter to arbitrarily small values  (e.g., \citealt{Naoz+12,Li+14,Hamers+16}).  The octupole-order Kozai-Lidov cycle takes place over a timescale given by \citep{Antognini15}
\begin{eqnarray}
&& \tau_{\rm KL}^{\rm oct} \sim \tau_{\rm KL}^{\rm quad}\sqrt{\frac{1-e_{\rm o}^{2}}{e_{\rm o}}\frac{a_{\rm o}}{a_{\rm p}}} \approx 5\times 10^8\,\,{\rm yr}\frac{(1-e_{\rm o}^{2})^{2}}{e_{\rm o}^{1/2}} \times \nonumber \\
&&\left(\frac{a_{\rm p}}{20\,{\rm AU}}\right)^{-2}\left(\frac{a_{\rm o}}{10^{3}{\rm AU}}\right)^{7/2}\left(\frac{M_{\rm o}}{0.4M_{\odot}}\right)^{-1}\left(\frac{M_{\star}}{1.43M_{\odot}}\right)^{1/2},
\label{eq:tauKL}
\end{eqnarray}
where we have calibrated the pre-factor of the analytic expression to the average value from our numerical calculations described below (Fig.~\ref{fig:KLtime}).  If the observed companion M dwarf of \tabby~is on an moderately eccentric bound orbit with a semimajor axis of $\sim 10^{3}$ AU similar to its observed separation, then $\tau_{\rm KL}^{\rm oct}$ is indeed comparable to the $\approx$ 2 Gyr lifetime of an F main sequence star.  Alternatively, the perturber could be an unobserved sub-stellar or even massive exoplanetary perturber, in which case the semi-major axis must be substantially smaller to give a Kozai-Lidov timescale of the appropriate order.

To more precisely constrain the required properties of the outer body, we perform orbital dynamics calculations of hierarchical triple systems using the N-body code IAS15, as included in the REBOUND Python package \citep{Rein&Liu12,Rein&Spiegel15}.  We employ a Monte Carlo approach of calculating the orbital evolution of a large number of systems in which we fix the properties of the (massless) exoplanet, but vary the semi-major axis and mass of the outer perturber across a 50$\times$50 logarithmic grid.  For each value of ($M_{\rm o}$, $a_{\rm o}$), we perform $N = 10^{4}$ calculations in which we sample the outer eccentricity, mutual inclination, and mutual longitude of pericenter from thermal, isotropic, and uniform distributions, respectively.  As a function of the exoplanet parameters, we determine the fraction of systems showing orbital flips (meaning the eccentricity passes through zero, likely indicating a strong interaction or collision with the central star) in a temporal range of $2\times 10^{8}$ yr $< \tau_{\rm KL}^{\rm oct} < 2\times 10^{9}$ yr, i.e. broadly consistent with the main-sequence lifetime of \tabby.  We remove systems from the sample for which the semi-major axis differences between the exoplanet and perturber are sufficiently small for Hills instability to operate \citep{Murray&Dermott99}.    

To sample the parameter space as densely as possible, we run our orbital simulations in the limit of Newtonian point particles.  While this approach has the advantage of shorter computational time, it neglects various short-range forces that may induce apsidal precession, and thereby suppress high-eccentricity excursions and orbit flips in some parts of parameter space.  The relevant short-range forces include (i) general relativistic corrections, (ii) tidal interactions, and (iii) the quadrupolar potential of \tabby{} due to rotational oblateness.  We follow \citet{Liu+15} in characterizing the importance of these different short-range forces, which induce apsidal precession rates of $\dot{\omega}_{\rm GR}$, $\dot{\omega}_{\rm tide}$, and $\dot{\omega}_{\rm rot}$, respectively.  The importance of these short-range forces is characterized by the following dimensionless ratios of precession rates:
\begin{align}
    \frac{\dot{\omega}_{\rm GR}}{\dot{\omega}_{\rm KL}}=&  \frac{3G(M_\star + M_{\rm p})^2 a_{\rm o}^3 (1-e_{\rm o}^2)^{3/2} }{a_{\rm p}^4 c^2 M_{\rm o}(1-e_{\rm p}^2)^{1/2} } \label{eq:SRF1} \\
    \frac{\dot{\omega}_{\rm tide}}{\dot{\omega}_{\rm KL}}=&  \frac{15 M_{\star}(M_\star + M_{\rm p})a_{\rm o}^3 (1-e_{\rm o}^2)^{3/2} k_{2, \rm p}R_{\rm p}^5}{a_{\rm p}^8 M_{\rm p} M_{\rm o}}\left( 1 + \frac{3e_{\rm p}^2}{2} + \frac{e_{\rm p}^4}{8}  \right) \label{eq:SRF2}  \\
        \frac{\dot{\omega}_{\rm rot}}{\dot{\omega}_{\rm KL}}=&  \frac{(M_\star + M_{\rm p}) a_{\rm o}^3 (1-e_{\rm o}^2)^{3/2} k_{\rm q,\star} \Omega_\star^2 R_\star^5}{G a_{\rm p}^5 M_\star M_{\rm o}(1-e_{\rm p}^2)^{3/2}}, \label{eq:SRF3}
\end{align}
where we have defined a ``characteristic'' KL precession frequency, 
\begin{equation}
    \dot{\omega}_{\rm KL} = \frac{1}{\tau^{\rm quad}_{\rm KL}(1-e_{\rm p}^2)^{1/2}}.
\end{equation}
Equation (\ref{eq:SRF2}) considers precession from the tidal bulge raised on the exoplanet by the star\footnote{One may also consider precession from the tidal bulge raised on the star by the exoplanet, by permuting $\{M_\star, M_{\rm p} \}$ in Eq.~\ref{eq:SRF2}, replacing $R_{\rm p} \to R_{\star}$, and also by using $k_{2, \star}$ in place of $k_{2, \rm p}$; in practice, however, the planetary bulge usually dominates the precession rate, so for the remainder of this paper we neglect precession from the bulge raised on the star.  }, and employs the exoplanetary (quadrupolar) tidal Love number $k_{2, \rm p}$.  Eq. \ref{eq:SRF3} considers precession from the quadrupolar moment created by the oblate surface of equilibrium for a rotating star\footnote{Likewise, one may use Eq.~\ref{eq:SRF3} to model precession from the spin-induced quadrupole moment of the exoplanet by permuting $\{M_\star, M_{\rm p} \}$, replacing $R_{\star} \to R_{\rm p}$, replacing $\Omega_{\star} \to \Omega_{\rm p}$, and also by using $k_{\rm q, \rm p} = k_{2, \rm p}/2$ in place of $k_{\rm q, \star}$.  We find that precession from the oblate quadrupole moment of \tabby{} dominates that from the oblate quadrupole moments of Neptunian exoplanets and is at worst comparable to that from Jovian exoplanets, so for simplicity we neglect exoplanetary spin for the remainder of this section (though it becomes of significant importance later, in \S \ref{sec:fate}).}, where $k_{\rm q, \star} = k_{2, \star} / 2$ is the stellar apsidal constant and $\Omega_\star = 2\pi / \mathcal{T}_\star$ the stellar rotational frequency.  If the high-eccentricity excursions of a Kozai cycle are terminated by one or more of the short-range forces listed above, the limiting, maximal eccentricity is given by solving the equation
\begin{equation}
    \frac{\dot{\omega}_{\rm GR}}{\dot{\omega}_{\rm KL}} + \frac{\dot{\omega}_{\rm tide}}{9\dot{\omega}_{\rm KL}} + \frac{\dot{\omega}_{\rm rot}}{3\dot{\omega}_{\rm KL}} = \frac{9}{8} \label{eq:arrest}
\end{equation}
for $e_{\rm p}$ \citep{Liu+15}.  We use this expression to analytically estimate in which regions of parameter space orbit flips are prevented by non-Newtonian precession.

Figure \ref{fig:KLtime} shows our results for the range of outer perturber properties ($M_{\rm o}, a_{\rm o}$) capable of driving a exoplanet into \tabby~throughout its main-sequence lifetime.  Different panels in the figure correspond to different values of the exoplanet semi-major axis (all cases assume a low initial exoplanet eccentricity $e_{\rm p} = 0.05$).  The allowed perturber properties overlaps with those of the putative M-dwarf companion of \tabby~\citep{Boyajian+16}, shown as a horizontal red line.  They also overlap those of many directly imaged exoplanet and brown dwarf companions\footnote{\url{http://exoplanet.eu}} (\citealt{Perryman11}), shown as red dots.  Red crosses show the properties of the perturbers explored in our fiducial numerical models of exomoon detachment described in $\S\ref{sec:detachment}$.  Our numerical results approximately match the analytically predicted power-law dependence $M_{\rm o} \propto a_{\rm o}^{7/2}$ for a fixed interval of $\tau_{\rm KL}^{\rm oct}$ from (Eq.~\ref{eq:tauKL}), as shown with a white line in Fig.~\ref{fig:KLtime}.

Analytic estimates (Eq. \ref{eq:arrest}, \citealt{Liu+15}) suggest that short-range forces, neglected in our simulations, may arrest KL cycles prior to exoplanet-star collision in a portion of parameter space.  For Jovian exoplanets, tidal interactions will prevent collisions for $a_{\rm p} \lesssim 10~{\rm AU}$, but collisions occur unimpeded for larger initial separations.  For Neptunian exoplanets, tides are less important than the spin-induced quadrupole moment of \tabby{}, which prevents collisions when $a_{\rm p} \lesssim 3 ~{\rm AU}$, but not for larger semimajor axes. In Fig. \ref{fig:KLtime}, and for the remainder of this paper, we have assumed a tidal Love number of $k_{2, \star}=10^{-2.356}$ of \tabby~(taken from Table 7 of \citealt{Claret&Gimenez92} for a $1.41M_\odot$ star of age $0.899~{\rm Gyr}$).  For Jupiter-like planets we take $k_{2, \rm p} = 0.535$, compatible with recent measurements by {\it Juno} \citep{Ni18}.  The Love number of Neptune has not been measured to notable precision, so we use a theoretical estimate, $k_{2, \rm p}=0.16$, for Neptune-like exoplanets \citep{Kramm+11}.

In summary, we conclude that a large fraction of the potential parameter space of exoplanet and outer perturber properties, potentially including those of the observed M-dwarf companion of \tabby, are sufficient  to driven the exoplanet into a close encounter during the main-sequence lifetime of \tabby.

\subsection{The Fate of Exomoons}
\label{sec:fate}

The properties of exomoons are essentially unconstrained observationally (however, see \citealt{Teachey&Kipping18} for a first candidate).  Nevertheless, the major gas giants in our Solar System are orbited by massive icy moons, such as the Galilean moons of Jupiter (Io, Europa, Ganymede, Callisto) and Titan orbiting around Saturn, some of the relevant properties of which we summarize in Table \ref{table:moons}.  Their masses are typically $\sim 10^{26}$ g, while their semi-major axes are $\approx 1-4$ percent of their exoplanet's Hill radii.  We now describe the process by which exomoons with similar properties around exoplanets can be removed by tidal forces and placed into stable orbits around the central star.

\subsubsection{Dynamics of Moon Detachment}
\label{sec:analytic}
We begin with analytic estimates of properties of exomoons after tidal detachment, before describing our detailed numerical study.  An exomoon orbiting its parent exoplanet with a semimajor axis $a_{\rm m}$ will be tidally detached from the exoplanet once $r_{\rm H} \lesssim a_{\rm m}$, where 
\begin{equation}
    r_{\rm H} \equiv q_{\rm p}\left(\frac{M_{\rm p}}{M_\star} \right)^{1/3}.
\end{equation}
is the Hill radius of the exoplanet at pericenter.  Though approximate, this detachment criterion serves as a useful benchmark.  

Upon separation, the exomoon acquires a new set of orbital elements, which we label as $\{\tilde{a}_{\rm m}, \tilde{e}_{\rm m}, \tilde{i}_{\rm m}, \tilde{\Omega}_{\rm m}, \tilde{\omega}_{\rm m}, \tilde{f}_{\rm } \}$.  Because detachment occurs at pericenter, we expect little change in specific orbital angular momentum of the exomoon, and thus an unchanged pericenter radius,
\be
\tilde{q}_{\rm m} \approx q_{\rm p}|_{\rm r_{\rm H} = a_{\rm m}} \approx a_{\rm p}\left(\frac{a_{\rm m}}{r_{\rm H,0}}\right) \approx 0.1{\rm AU} \left(\frac{a_{\rm p}}{10{\rm AU}}\right)\left(\frac{a_{\rm m}}{0.01r_{\rm H,0}}\right),
\label{eq:qm}
\ee  
where we have normalized the initial semi-major axis of the exomoon to the exoplanet's initial Hill radius $r_{\rm H,0} \approx a_{\rm p}(M_{\rm p}/M_{\star})^{1/3}$ for an approximately circular initial exoplanet orbit.

The process of tidal disruption imparts energy to the exomoon.  Roughly speaking, the detached exomoon's new specific orbital energy will be drawn from a distribution of half-width \citep{Rees88}
\begin{equation}
    \Delta \epsilon \simeq \frac{GM_{\rm p}}{a_{\rm m}} \left( \frac{M_\star}{M_{\rm p}} \right)^{1/3}.
\end{equation}
This energy spread represents a Taylor expansion of the star's gravitational potential around the initial,  exoplanetocentric orbit of the exomoon.  The semi-major axis of the detached exomoon depends on this energy spread relative to the initial energy of the exoplanet's orbit $\epsilon_{\rm p} \equiv GM_\star /(2a_{\rm p})$,
\be
\frac{\Delta \epsilon}{\epsilon_{\rm p}} \approx \frac{2a_{\rm p}}{a_{\rm m}} \left( \frac{M_{\rm p}}{M_{\star}} \right)^{2/3} \approx 20\left(\frac{a_{\rm m}}{0.01r_{\rm H,0}}\right)^{-1}\left(\frac{M_{\rm p}}{10^{-3}M_{\star}}\right)^{1/3},
\label{eq:deltaE}
\ee
If $\Delta \epsilon \ll \epsilon_{\rm p}$, then the detached exomoon inherits the initial semimajor axis of the exoplanet, i.e. $\tilde{a}_{\rm m} \approx a_{\rm p}$.  Conversely, if $\Delta \epsilon \gg \epsilon_{\rm p}$, then the exomoon has a roughly $50\%$ probability of immediate ejection; if the exomoon stays bound, it will typically have a semimajor axis 
\be \tilde{a}_{\rm m} \approx \frac{GM_\star}{2\Delta \epsilon} = \frac{a_{\rm m}}{2}\left(\frac{M_{\star}}{M_{\rm p}}\right)^{2/3} =a_{\rm p}\left(\frac{\epsilon_{\rm p}}{\Delta \epsilon}\right) \ll a_{\rm p}.
\label{eq:amtilde}
\ee  Equation (\ref{eq:deltaE}) shows that for gas giant exoplanets or larger, and exomoon semi-major axes $a_{\rm m} \lesssim 0.04r_{\rm H}$ similar to those of the major moons in our Solar System (Table~\ref{table:moons}), we are in the $\Delta \epsilon \gg \epsilon_{\rm p}$ limit.  The orbital period of the exomoon after detachment is given in this limit by
\begin{eqnarray}
\tilde{t}_{\rm orb} &=& 2\pi \left(\frac{\tilde{a}_{\rm m}^{3}}{GM_{\star}}\right)^{1/2} \approx \pi \left(\frac{a_{\rm m}^{3}}{GM_{\star}}\right)^{1/2}\left(\frac{M_{\star}}{M_p}\right)  \\
&\approx& 4.2\,{\rm yr}\,\left(\frac{M_p}{10^{-4}M_{\star}}\right)^{-1}\left(\frac{a_{\rm m}}{0.01{\rm AU}}\right)^{3/2}\left(\frac{M_{\star}}{1.4M_{\odot}}\right)^{-1/2}, \notag
\end{eqnarray}
where $a_{\rm m}$ is normalized to a characteristic value for major moons in our Solar System (Table \ref{table:moons}).

\begin{figure}
\includegraphics[width=0.46\textwidth]{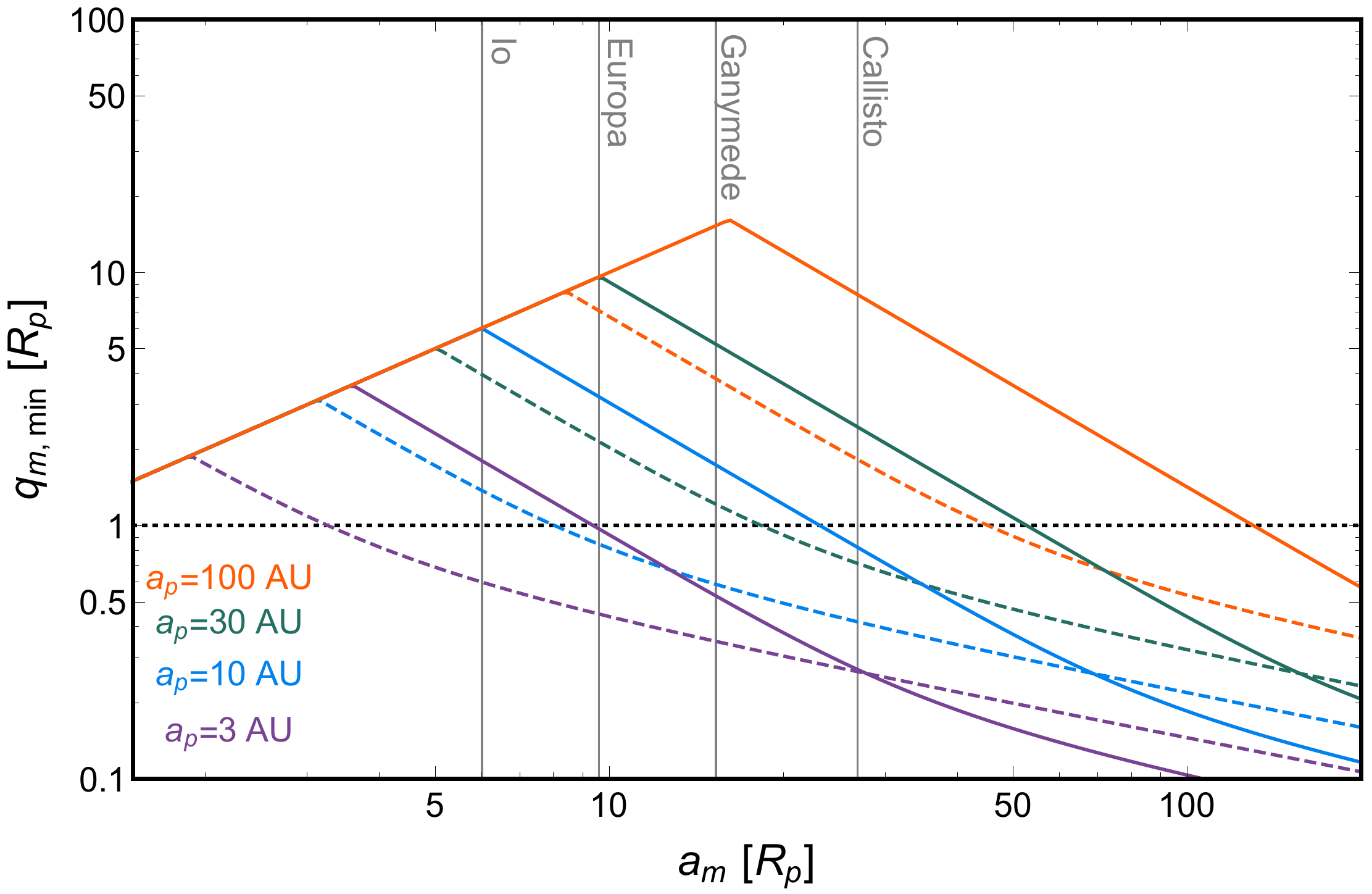}

\caption{A sufficiently wide initial exomoon orbit (large $a_{\rm m}$) could drive the exomoon into a collision with its host planet via ``secondary KL cycles'' driven by secular torques from the central star.  Here we show the minimum achievable exomoon pericenter $q_{\rm m, min}$ through the secondary KL cycle as a function of $a_{\rm m}$ (each normalized to planet radius, $R_{\rm p}$), which are calculated by solving equation \ref{eq:arrest} at the time of detachment ($q_{\rm p}=r_{\rm H}$).   Solid lines correspond to Jupiter-type exoplanets, and dashed lines correspond to Neptune-type exoplanets; the orange, green, blue, and purple curves indicate exoplanetary semimajor axes $a_{\rm p}$ of 100~{\rm AU}, 30~{\rm AU}, 10~{\rm AU}, and 3~{\rm AU}, respectively.  For $q_{\rm m, min}/R_{\rm p} \lesssim 1$ (black dotted line), a moon-planet collision is possible (though not guaranteed).  For comparison, vertical gray lines show the semimajor axes of the Galilean satellites, all of which have $q_{\rm min,min} \gtrsim R_{\rm p}$ and therefore would not be at risk of a secondary KL-driven collision.  }
\label{fig:secondaryKL}
\end{figure}

Thus far we have implicitly assumed that the exomoon survives up to the point of tidal detachment ($q_{\rm p} \approx r_{\rm H}$).  However, one might be concerned that prior to this point KL cycles {\it between the exomoon and the exoplanet}, with \tabby~acting as the hierarchical tertiary, could eliminate the exomoon (via a collision with its parent exoplanet).  Although the secular dynamics of hierarchical quadruples have been investigated in greater detail elsewhere \citep{Munoz&Lai15, Hamers+15, Hamers&Lai17}, we are mainly concerned with evaluating the importance of secular eccentricity oscillations in the pre-detachment exomoon-exoplanet system, which we refer to as ``secondary KL cycles.''  We evaluate the risk of a pre-detachment exomoon-exoplanet collision in a secondary KL cycle by permuting the variables of Eq. \ref{eq:arrest}.  We find that GR is entirely negligible for the orbit of the inner exomoon/exoplanet binary.  The tidal bulge raised by the exomoon on the exoplanet is at best marginally effective at preventing collisions induced by secondary KL cycles.  However, the spin-induced quadrupole moment of the exoplanet will be extremely effective at preventing these collisions, provided that exoplanetary gas giants rotate at $\gtrsim 10\%$ of breakup (as do all Solar System gas giants).  If we neglect tides and GR, we find that in order to avoid a collision due to secondary KL, the initial semimajor axis of the exomoon must be less than a critical value given by
\begin{equation}
    \frac{a_{\rm m}}{R_{\rm p}} < \frac{2k_{2, \rm p}}{3\sqrt{3}} \frac{\Omega_{\rm p}}{GM_{\rm p}/R_{\rm p}^3} \left(\frac{M_{\rm p}}{M_\star} \right)^{1/4} \left(\frac{a_{\rm p}}{R_{\rm p}} \right)^{3/4}. \label{eq:secondarySafety}
\end{equation}
This criterion implies that the most loosely bound exomoons will be vulnerable to secondary KL (although whether or not they have time to collide with their parent exoplanet depends on other parameters of the problem).  However, note that Eq. \ref{eq:secondarySafety} is a sufficient but not necessary criterion for exomoon survival.  An exomoon/exoplanet system could fail to satisfy Eq. \ref{eq:secondarySafety}, but nonetheless survive to the point of tidal detachment because either (i) the mutual inclination of the inner triple is unfavorable, or (ii) the octupole-order secondary KL cycles do not have enough time to reach their eccentricity maxima.  

Figure \ref{fig:secondaryKL} illustrates the parameter space of secondary KL cycles more precisely by showing the minimum achievable exomoon pericenter as a function of the exomoon semi-major axis $a_{\rm m}$, which we have obtained by solving Eq. \ref{eq:arrest} numerically to account for precession from both a tidal bulge and rotational oblateness.  For the most relevant region of parameter space ($a_{\rm p} \gtrsim 10~{\rm AU}$), the maximum ``safe'' $a_{\rm m}$ is between $10 R_{\rm p}$ and $100R_{\rm p}$, a range that contains the semimajor axes of many icy moons in the Solar System.  For this reason, we neglect secondary KL for the remainder of this paper, but caution that it may be important for exomoons on very wide orbits.

\begin{table*}
\caption{Some properties of major moons in the Solar System \label{table:moons}}
\begin{tabular}{cccccccc}
\hline\hline
Moon & $m_{\rm m}$ & $R_{\rm m}$  & $a_{\rm m}$ & $a_{\rm m}$ & $\alpha_{\rm m}^{(1)}$  & ${\tilde{a}_{\rm m}}^{(2)}$ & $\langle t_{\rm sub} \rangle^{(3)}$ \\
\hline
- &  ($10^{26}$ g) &  ($10^{8}$ cm) & (AU) & ($r_H$) & - & (AU) & (10$^{4}$ yr) \\
\hline
\hline
Io       & 0.89 & 1.8 & 2.81$\times 10^{-3}$ & 8.31$\times 10^{-3}$ & 0.63 & 0.18 & 0.62 \\
Europa   & 0.48 & 1.6 & 4.49$\times 10^{-3}$ & 1.33$\times 10^{-3}$ & 0.67 & 0.28 & 1.58 \\
Ganymede & 1.49 & 2.6 & 7.16$\times 10^{-3}$ & 2.12$\times 10^{-2}$ & 0.43 & 0.45 & 3.99 \\
Callisto & 1.08 & 2.4 & 1.26$\times 10^{-2}$ & 3.73$\times 10^{-2}$ & 0.22 & 0.78 & 12.2 \\
Titan    & 1.35 & 2.6 & 8.17$\times 10^{-3}$ & 1.98$\times 10^{-2}$ & 0.22 & 0.51 & 5.18 \\
\hline\\
\end{tabular}

$^{(1)}$Albedo, $^{(2)}$Estimate of the semi-major axis following detachment into an orbit around a star with properties similar to\,\tabby\, (Eq.~\ref{eq:amtilde}), $^{(3)}$Orbit averaged minimum sublimation time after detachment (Eq.~\ref{eq:tsubElliptic}), estimated using equation (\ref{eq:qm}) for $\tilde{q}_{\rm m}$. 
\end{table*}

\begin{figure}
\includegraphics[width=0.47\textwidth]{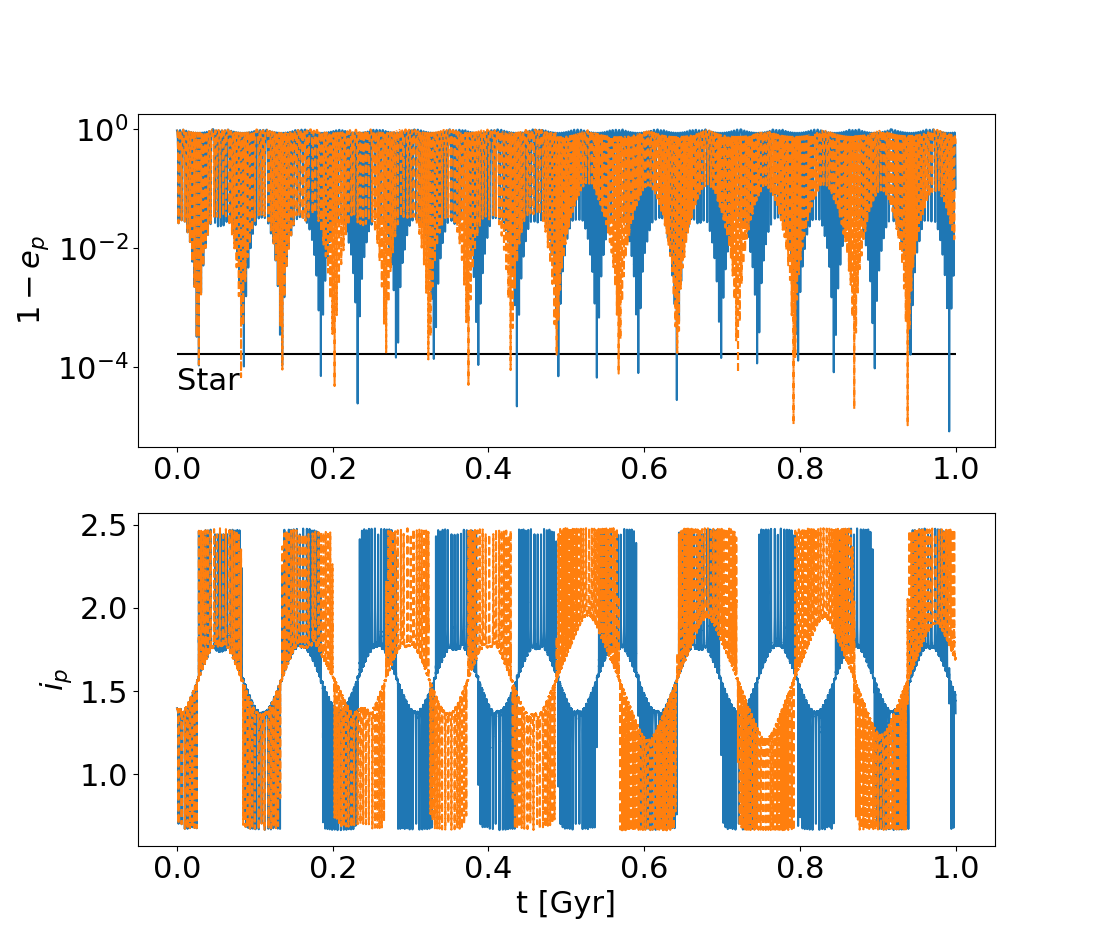}

\caption{Time evolution of the dimensionless angular momentum deficit $1 - e_{\rm p}$ (top panel) and inclination $i_{\rm p}$ (bottom panel) of the exoplanet's orbit for a representative example: the initial exoplanet semi-major axis is $a_{\rm p} = 45$ AU, and the outer perturber has $M_{\rm o} = 0.4 M_{\odot}$, $a_{\rm o} = 900$ AU, and $e_{\rm o} = 0.45$.  Solid blue lines show the calculation in Newtonian gravity, while dashed orange lines show the results when relativistic effects are included at 1 PN order.  Results are qualitatively similar between the two cases.}
\label{fig:GR}
\end{figure}

\begin{figure}
\includegraphics[width=0.43\textwidth]{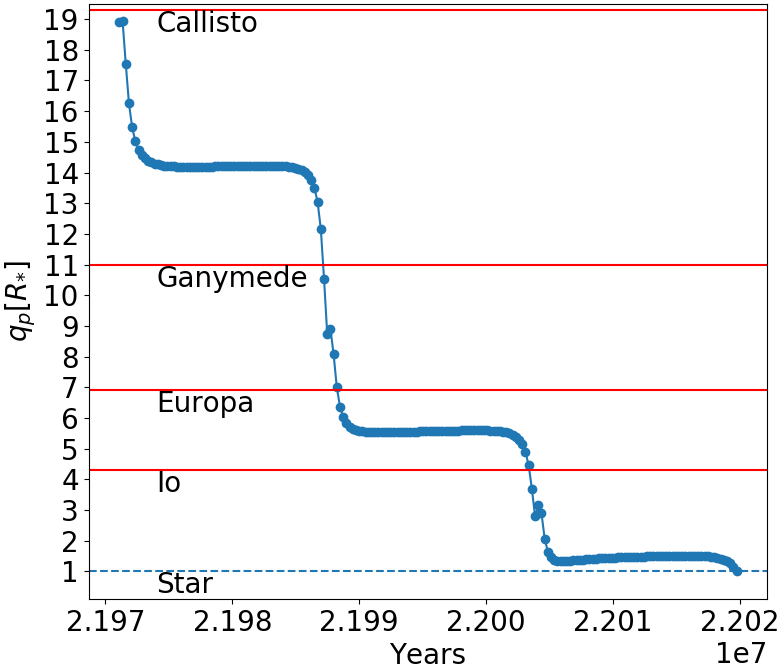}
\includegraphics[width=0.43\textwidth]{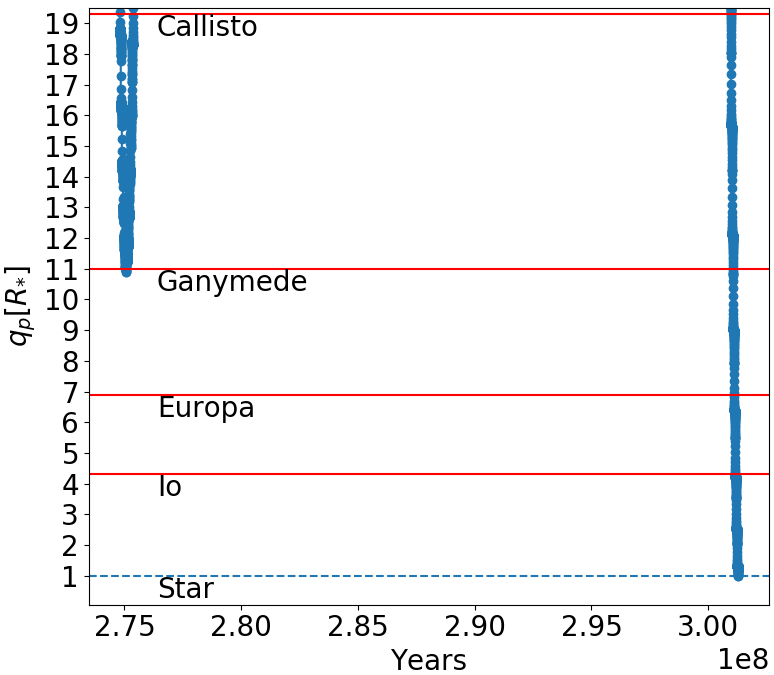}
\includegraphics[width=0.43\textwidth]{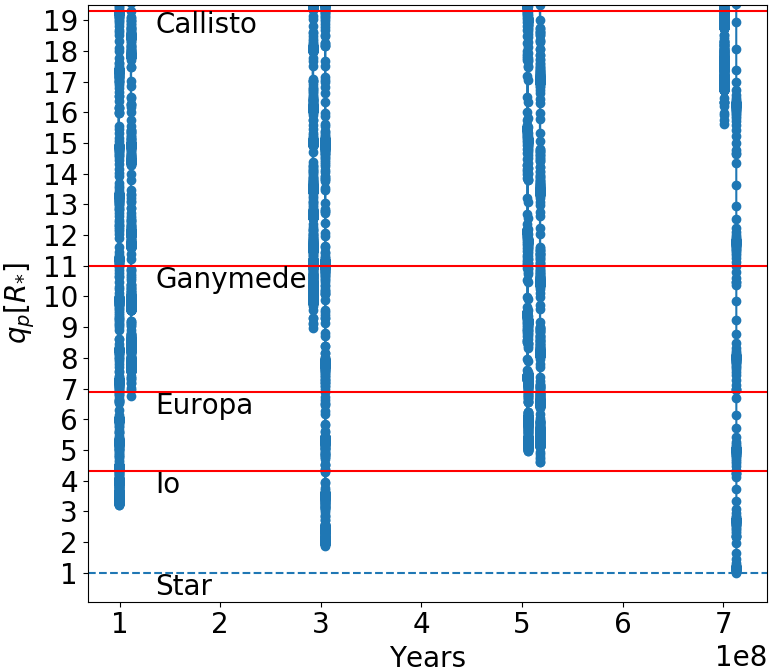}
\caption{Examples of the time evolution of the exoplanet pericenter radius $q_{\rm p}$ (blue points, in units of the stellar radius), prior to the exoplanet's destruction.  For comparison we show the semi-major axes ($\sim$ tidal stripping radii) of the Galilean moons with horizontal solid lines.  Cases shown include those characterized by: (a) punctuated episodes of rapid eccentricity growth that slowly strip off all the exomoons; (b) first passage that strips off the outer exomoons but the exoplanet-star collision occurs on the second flip several hundred million years later; (c) The ultimate collision happens only on the fourth flip, after several episodes closely approach the stellar surface, in which case all the exomoons are tidally stripped long before the exoplanet is destroyed.}
\label{fig:collision}
\end{figure}

\subsubsection{N-Body Simulations}
\label{sec:detachment}

To explore in greater detail the conditions giving rise to exoplanet-star collisions, we first evolve a range of three-body systems (i.e. excluding the exomoon), again using the N-body code IAS15.  Unlike in \S \ref{sec:Kozai}, we now make use of the accompanying REBOUNDx package, which includes 1PN relativistic effects (Rein et al., in prep).  

Our fiducial scenario considers a Jupiter-like exoplanet ($M_{\rm p} = 10^{-3}M_{\odot}$) with a range of initial orbital elements ($a_{\rm p} =\{12,30,45\} {\rm AU}$, $e_{\rm p} = 0.05$, $i_{\rm p} = 1.22$).  We consider properties for the outer perturber consistent with the potential M-star companion of \tabby~ ($M_{\rm o} = 0.4M_{\odot}$, $a_{\rm o} = 900$ AU, $e_{\rm o} = \{0,0.5,0.9\}$), in addition to several cases corresponding to a less massive, and currently unobserved, sub-stellar companion ($M_{\rm o} = \{0.01,0.05,0.08\}M_{\odot}$, $a_{\rm o} = \{500,750,1000,1250,1500\}$ AU, $e_{\rm o} = 0.7071$).
\begin{table*}
\caption{Properties of Simulations and Results \label{table:simresults}}
%\centering
\begin{tabular}{ccccccccccc}
\hline\hline
Model & $M_{\rm p}$ & $M_{\rm o}$ & $i_{m,0}$ & $f_{\rm ej}^{(1)}$ & $f_{\rm coll}^{(2)}$ & $f_{\rm surv}^{(3)}$ & $\langle{q_{\rm m}}\rangle^{(4)}$ & $\langle{e_{\rm m}}\rangle^{(4)}$ & $M(\langle{t_{\rm sub}\rangle})^{(5)}$ & $M(t_{\rm solid})^{(6)}$\\
\hline
- & ($10^{-3}$ $M_{\odot}$) & ($10^{-2}$ $M_{\odot}$) & - & - & - & - & (AU) & - & (yr) & (yr) \\
\hline
Fiducial    & 1 & 5 & 70$^{\circ}$   & 0.524 & 0.398 & 0.078  & 0.08$_{0.02}^{2.09}$ & 0.84$_{0.55}^{0.94}$ & $1.38 \times 10^5$ & $1.92 \times 10^5$\\
Small exoplanet  & 0.1  & 5 & 70$^{\circ}$   & 0.566    & 0.324 & 0.11  & 1.33$_{0.064}^{9.21}$ & 0.84$_{0.31}^{0.96}$ & $6.33 \times 10^6$ & $1.72 \times 10^5$\\
%Small Perturber   & 1  & 0.5 & 70$^{\circ}$   & x    & x & x & x & ? & ? & ?\\
%Other 1     & 1     & x     & 70$^{\circ}$   & x    & x & 0.63  & 0.2 & ? & ? & ?\\
%Other 2     & 1     & x     & 70$^{\circ}$   & x    & x & 1.8   & 0.2 & ? & ? & ?\\
\hline\\
\end{tabular}

$^{(1)}$Fraction of exomoons ejected from the system. $^{(2)}$Fraction of exomoons destroyed by a collision with the central star. $^{(3)}$Fraction of surviving exomoons after exoplanet destruction. $^{(4)}$Subscripts and superscripts represent the 10th and 90th quantiles of the data set, respectively. $^{(5)}$Median minimum sublimation time of the detached exomoons (Eq.~\ref{eq:tsubElliptic}) for a star with properties similar to \tabby.   $^{(6)}$Median maximum timescale for accretion of solid debris through the disk by Poynting-Robertson drag (Eq.~\ref{eq:tsolid}).
\end{table*}

We initialize each permutation of the initial conditions and evolve the systems, in pairs of otherwise identical runs that include or neglect relativistic corrections.  For the parameter space we survey, we find that relativistic effects generally do not significantly alter the collision timescale of the exoplanet (see Fig.~\ref{fig:GR} for a representative example).  We also examine the evolution of the separation between the primary and the exoplanet during their closest approaches ($r<20 R_\star$) in order to quantify the number of such approaches prior to the final destructive encounter.  A few examples are shown in Figure \ref{fig:collision}.  In most cases, such as panels (a) and (b), the majority of the Galilean satellites would not be tidally detached from the exoplanet during the eccentricity maximum of a {\it non-terminal} KL cycle.  In other words, the exoplanet would retain most of its population of (Galilean-like) exomoons until the final, terminal KL cycle.  However, in cases like (c), most exomoons are detached in brief excursions to high eccentricity which occur many such KL cycles before the exoplanet's final destruction.  Although in principle, an exomoon detached after the first close passage could again be subject to scattering by the exoplanet once the latter finally returns, in practice the timescale $\gtrsim 10^{8}$ yr between close passages greatly exceeds the estimated lifetime of the detached exomoon due to photo-evaporation ($\S\ref{sec:outgassing}$).

Our suite of simulations provides several promising sets of initial conditions for subsequent exomoon detachment, with collisions times appropriate to the estimated age of \tabby~(i.e. those within the allowed band in Fig.~\ref{fig:KLtime}).  We then consider the full four-body system by adding exomoons to a subset of the N-body simulations once the exoplanet's pericenter becomes less than $3.5 r_{H}$, where the tidal radius is defined as above with respect to the initial semi-major axis of the exomoon.  As the fiducial model, we take parameters $M_{\rm p} = 10^{-3}M_{\odot}, a_{\rm p}=30$ AU, $M_{\rm o} = 0.05 M_{\odot}$, $a_{\rm o} = 500$ AU, and $e_{\rm o} = 0.7071$ from the prior set of runs.  
%This case was chosen mainly for computational efficiency, as it gave a relatively short exoplanet-star collision time.  
We perform a set of $N = 1000$ simulations, in each case sampling the initial exomoon orbits from a log uniform distribution of initial semimajor axes which range from twice the radius of Jupiter to 4.7 times the semi-major axis of Callisto.   Our set of models and the results are summarized in Table \ref{table:simresults}.  For the fiducial model we find that 8\% of the exomoons survive in a bound orbit after the exoplanet encounters the star.  Of the remainder, 40\% of the exomoons are themselves destroyed through a stellar collision, while 52\% experience strong kicks and are gravitationally unbound from the stellar system. 

\begin{figure*}
\includegraphics[width=1.0\textwidth]{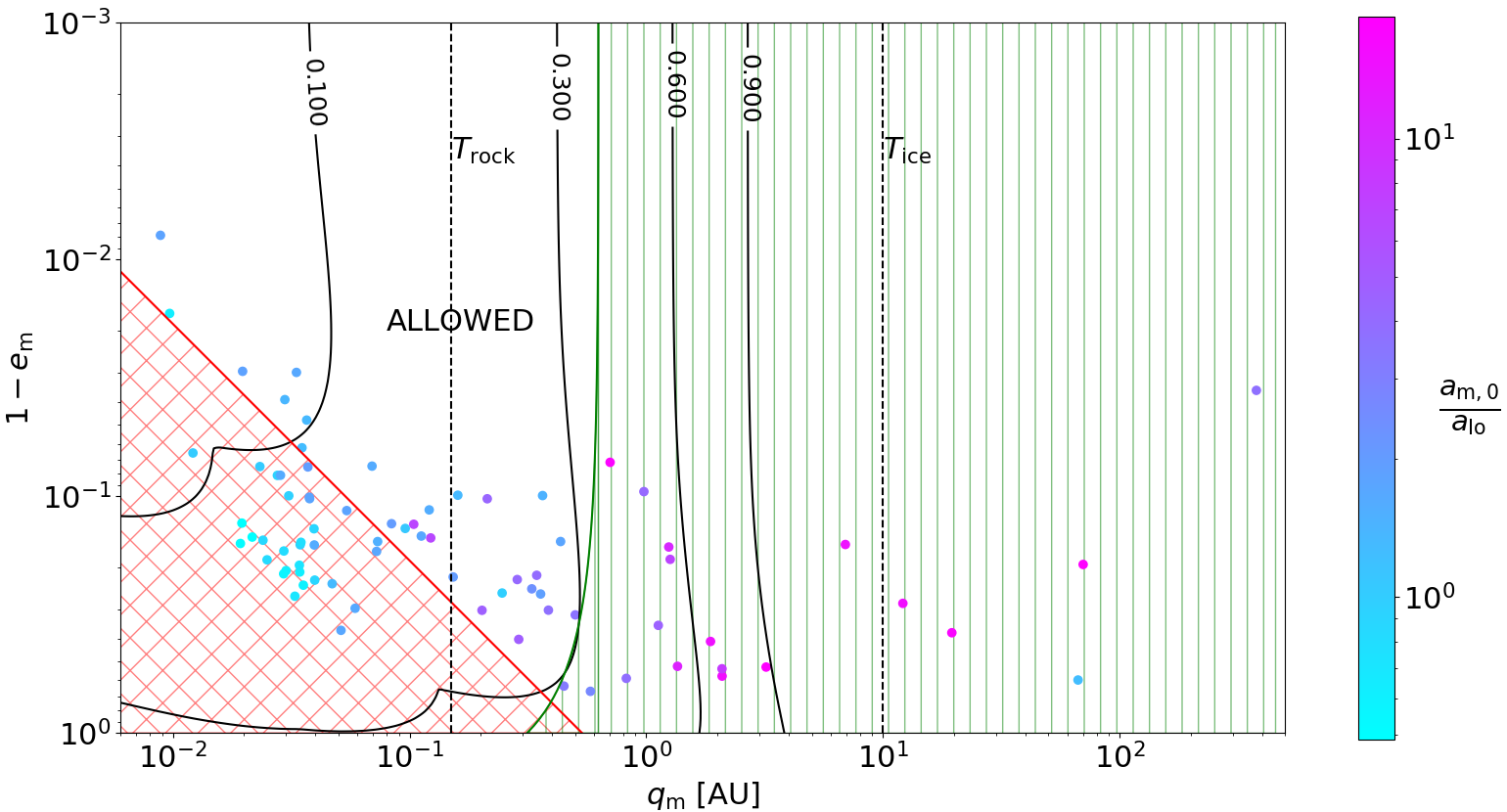}

\caption{Comparison of the orbital properties of the tidally detached exomoons from our N-body simulations to observational constraints on those of the outgassing bodies orbiting \tabby.  Colored dots show the final orbital properties of tidally detached exomoons from our fiducial model that survived detachment on stable elliptical orbits, with the color indicating their initial semi-major axis around the parent exoplanet.  For comparison, we show constraints from \citet[][their Fig.~4]{Schaefer+18} on the properties of the debris responsible for generating the light curve dips in \tabby.  The red area is ruled out by the periodicity constraint $t_{\rm orb} > 750$ days, while the green area is ruled out by the requirement to produce dips as short as 0.4 days (as observed); as a result, only the white portion of parameter space is observationally permitted.  Black contours represent the fractional dimming produced by dust with those orbital elements, assumed to be composed of astrosilicates with a size distribution down to 0.1 micron. }
\label{fig:dust}
\end{figure*}

Figure \ref{fig:dust} shows the final orbital parameters of the surviving detached exomoons in our the fiducial model. For comparison, we show observational limits on the orbital parameters of the dusty debris clouds responsible for dimming \tabby~based on the duration of the observed dips and their lack of periodicity, adapted from a similar figure in \citet{Schaefer+18}.  Although many of the simulated exomoons have orbital periods which lie in the region forbidden by periodicity constraints from Kepler, a substantial fraction fall within the allowed region.  Furthermore, almost all of the detached bodies reside on orbits inside the ice sublimation line (shown as a vertical dashed line in Fig.~\ref{fig:dust}), permitting subsequent outgassing.  We conclude that the tidal detachment of exomoons with initial properties similar to those of major moons in our own Solar System is broadly consistent with those in orbit around \tabby.

\begin{figure}
\includegraphics[width=0.5\textwidth]{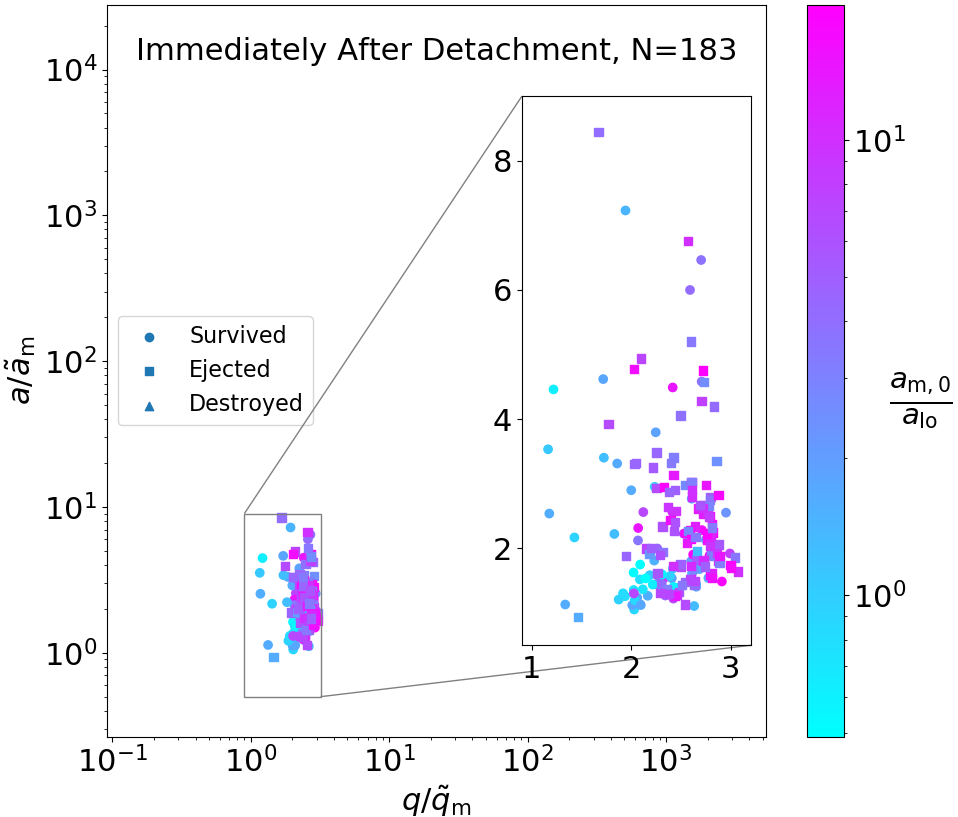}
\includegraphics[width=0.5\textwidth]{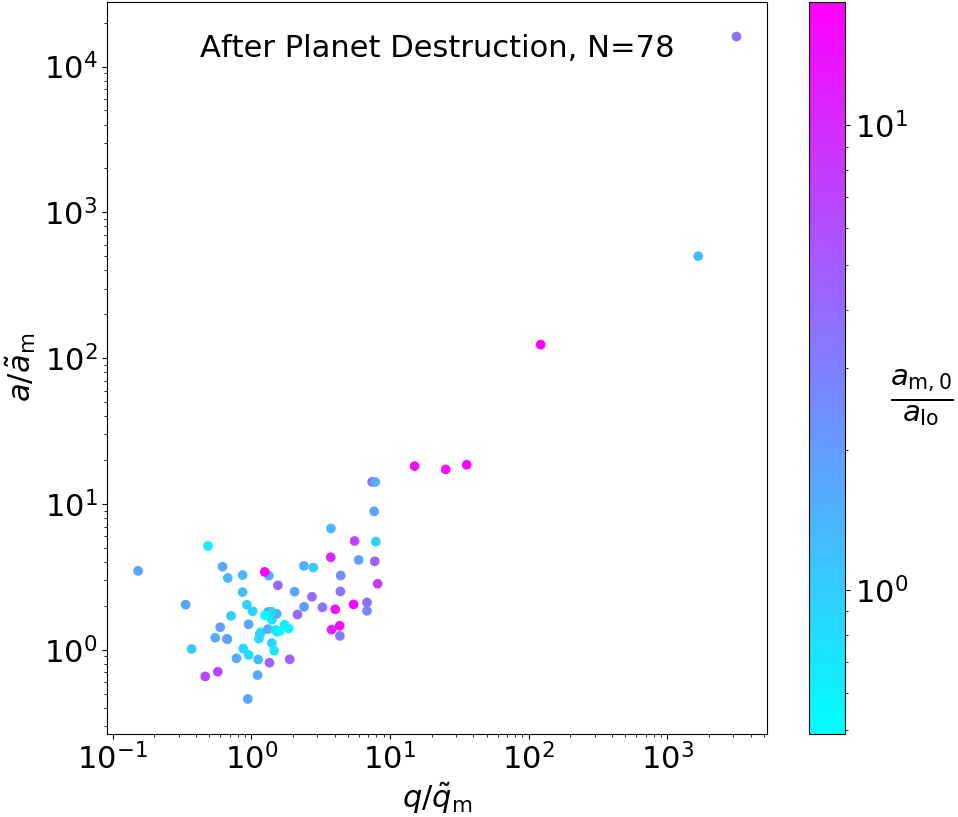}

\caption{Post-detachment semimajor axes $a_{\rm m}$ and pericenter radii $q_{\rm m}$ of the sample of tidally-detached exomoons from our fiducial model, normalized to the analytically predicted values (Eqs. \ref{eq:qm}, \ref{eq:amtilde}).  The top panel shows the distribution at the first output timestep following detachment (while the exoplanet is still present), while the bottom panel shows the distribution after the exoplanet has been destroyed.  The change in orbital properties from the top to bottom panel results from ongoing gravitational perturbations from the exoplanet, even after the exomoon has detached.}
\label{fig:orbitproperties}
\end{figure}

Figure \ref{fig:orbitproperties} shows the same orbital parameters as in Figure \ref{fig:dust}, but now normalized to our analytic estimates for the mean pericenter radius (Eq.~\ref{eq:qm}) and semi-major axis (Eq.~\ref{eq:amtilde}).  The top panel shows the distribution soon after tidal detachment, while the bottom panel shows the final values after the exoplanet has been destroyed.  The distribution of $q_{\rm m}$ in the top panel indeed clusters around $\tilde{q}_{\rm m}$ in the expected manner \citep{Rossi+17}.  The finite width of the distribution of semi-major axes $\tilde{a}_{\rm m}$ results, in part, from the energy spread imprinted by the process of tidal disruption \citep{Rees88}.  The widths of the distributions of post-detachment orbital elements also grow significantly going from the top to the bottom panel, demonstrating the sizable influence of ongoing gravitational perturbations from the exoplanet on the exomoons prior to the exoplanet's destruction.  Indeed, even some of the "initial" spread shown in the top panel arises because we have sampled the exomoon orbital elements only after the first readout time-step of the simulation following detachment, while gravitational perturbations by the exoplanet can begin immediately after tidal detachment.  

Beyond our fiducial model, we have explored separate runs varying the mass of the exoplanet or perturber.  If we reduce the mass of the perturber by a factor of 10 from the fiducial model to $M_{\rm o} = 10^{-4}M_{\odot}$, the surviving exomoon fraction remains similar ($\sim 11\%$) but the median pericenter radius after detachment increases substantially.

\begin{figure}
\includegraphics[width=0.5\textwidth]{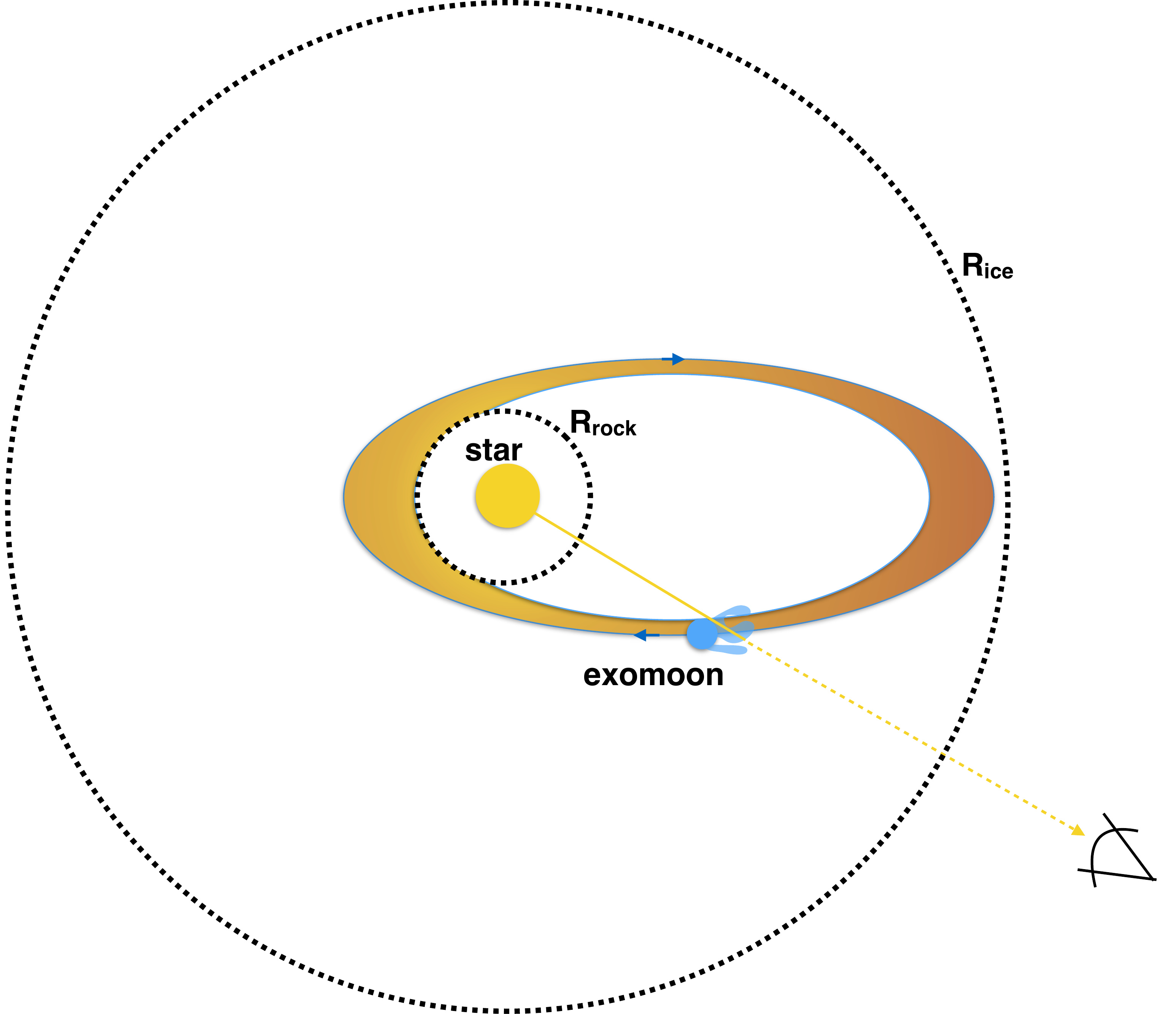}
\caption{Schematic illustration of our scenario for explaining short- and long-timescale dimming of the stellar light in systems like \tabby.  Gravitational perturbations from a massive outer body (e.g. through the Kozai-Lidov mechanism) pump up the eccentricity of an exomoon-bearing exoplanet and drive it into a destructive encounter with the central star.  Prior to the destruction of the exoplanet, the exomoons orbiting it are detached from the exoplanet by tidal forces and a fraction are placed onto new, highly-eccentric orbits about the star.  The volatile-rich exomoons now find themselves orbiting inside the ice sublimation radius $R_{\rm ice}$, leading to strong outgassing of gaseous and solid materials which shroud the exomoon in an opaque debris cloud.  For an observer situated in the exomoon orbital plane, this debris tail (which includes small particles capable of generating achromatic extinction) generates deep dips in the lightcurve when the exomoon transits the stellar disk.  While small dust grains are quickly removed from the system by radiation pressure, larger solid particles inherit the initial eccentric orbit of the exomoon.  After phase-mixing over a period of up to several orbits, these larger solid particles accrete more slowly through Poynting-Robertson drag, giving rise to secular timescale changes in the light curve as the radial optical depth through the disk midplane slowly evolves.  Full evaporation of the moon, and thus the duration of the dipping/secular dimming behavior, typically takes place over $\sim 10^{4}-10^{6}$ years, depending on the properties of the exomoon and its orbit about the star.}
\label{fig:cartoon}
\end{figure}

\section{Evaporation, Outgassing, and Obscuration of the Star}
\label{sec:outgassing}
As the new orbits of the tidally-detached exomoons typically bring them inside the ice line (Fig.~\ref{fig:dust}), they will experience strong stellar irradiation that evaporates volatile material from their surfaces.  In a scaled-up analogy to a comet passing near the Sun, the resulting outgassing of gas and solid particles (dust) may be powerful enough to create an opaque cloud around the exomoons, which when transiting the star could produce deep minima ("dips") in the light curve similar to those observed from \tabby~(see Fig.~\ref{fig:cartoon} for a cartoon).  Large dust particles released during this process will, over many periods, fill out the orbit of the exomoon and could provide a source of long-term secular evolution in the stellar light curve \citep{Wyatt+18}, as the dust disk is intermittently fed from time-variable outgassing episodes, and accretes slowly through Poynting-Robertson drag.  In this section we discuss the process of evaporation, the evolution of the solid debris, and its implications for the survival time of the exomoon. 
\subsection{Evaporation and Outgassing}
At distances $r \gg R_{\star}$ from the star, the equilibrium temperature of a body of albedo $\alpha_{\rm m}$ is given by 
\begin{equation}
    T_{\rm eq} = \left( \frac{L_\star (1-\alpha_{\rm m})}{4\pi \sigma r^2}\right)^{1/4} \approx T_{\star}(1-\alpha_{\rm m})^{1/4}\left(\frac{r}{R_{\star}}\right)^{-1/2} ,
\label{eq:Teq}
\end{equation}
where $T_{\star}$ is the stellar effective temperature ($T_{\star} \approx 6750$ K for \tabby).  Unless the initial exomoon system begins much more weakly bound to its parent exoplanet than are the major moons of our Solar System, then the pericenter distances of tidally separated exomoons (Eq.~\ref{eq:qm}; Fig.~\ref{fig:dust}) will fall well inside the water ice line ($T_{\rm eq} \gtrsim T_{\rm ice} \approx 170$ K, corresponding to $R_{\rm ice} \sim 10$ AU for \tabby), and in some cases also within the silicate sublimation line ($T_{\rm rock} \approx 1500$ K, corresponding to $R_{\rm rock} \approx 0.15$ AU).  Detached exomoons with volatile-rich surface layers will experience massive outgassing inside the ice line more analogous to comet disintegration than to atmospheric Jeans escape.  For example, a detached exomoon with $\tilde{q}_{\rm m} = 0.3~{\rm AU}$, $\tilde{e}_{\rm m} = 0.5$ (see Fig. \ref{fig:dust}), and $\alpha_{\rm m}=0.5$ will have an equilibrium surface temperature varying between apocenter and pericenter as $408~{\rm K} < T_{\rm eq} < 707~{\rm K}$.  The corresponding mean thermal speeds for gas-phase ammonia and water range between $0.61~{\rm km~s}^{-1} < v_{\rm th} < 0.80~{\rm km~s}^{-1}$, which are large fractions of typical moon escape velocities in the Solar System (e.g. $v_{\rm esc} \approx 2~{\rm km~s}^{-1}$ for the Galilean satellites).  While mass loss rates will thus be high, the detached exomoons will not evaporate within a single pericenter passage, but instead require many orbits to completely sublimate their volatile content.  

\begin{figure}
\includegraphics[width=0.5\textwidth]{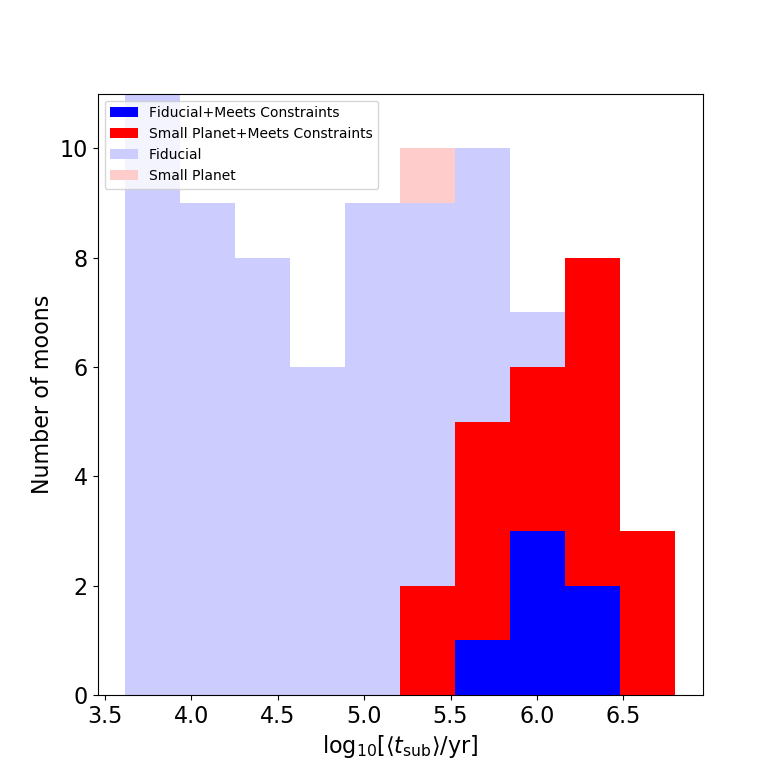}
\includegraphics[width=0.5\textwidth]{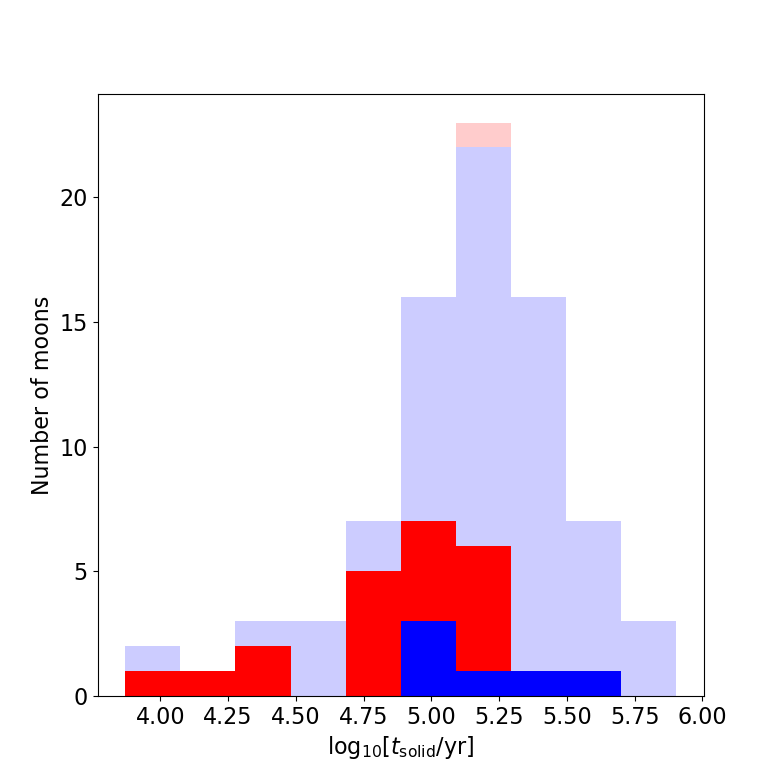}

\caption{Histograms showing the distribution of timescales among the population of surviving, detached exomoons in our Fiducial and Small Exoplanet models.  The results are broken down separately by (i) model and (ii) whether or not the resulting orbits satisfy the observational constraints on \tabby~from \citet{Schaefer+18} (the allowed region in Fig.~\ref{fig:dust}); colors indicate different subsets of results.  {\it Top:} Energy-limited sublimation time for the surviving exomoons from our fiducial model, computed using Eq.~\ref{eq:tsubElliptic} for an assumed exomoon mass $M_{\rm m} \approx 10^{26}$ g.  We note that this is an absolute minimum evaporation time.  {\it Bottom:} Draining timescale of the solid particle disk due to PR drag (Eq.~\ref{eq:tsolid}).}
\label{fig:histogram}
\end{figure}

Consider a toy two-component model for an exomoon, with mean density $\rho_{\rm m}$ and a fraction $f_{\rm vol}$ of its total mass in volatile form with latent heat of sublimation $Q_{\rm vol} \approx 3 \times 10^{10}~{\rm erg~g}^{-1}$.  At fixed separation $r$, the absolute minimum time for evaporation is the time needed to absorb the energy that would sublimate and boil off\footnote{For small, gravity-dominated exomoons comparable in size to the Galilean satellites, $Q_{\rm vol} \sim v_{\rm esc}^2/2$, but for much larger bodies ($v_{\rm esc} \gtrsim 10~{\rm km~s}^{-1}$), $Q_{\rm vol}$ will be highly subdominant.  Conversely, exomoons too small to be in hydrostatic equilibrium will see Eq. \ref{eq:tsubSimple} dominated by $Q_{\rm vol}$.} the entire volatile content (i.e. we optimistically assume immediate thermal escape and neglect radiative losses), which is
\begin{equation}
    t_{\rm sub} = \frac{16\pi}{3} \frac{(Q_{\rm vol}+ v_{\rm esc}^2/2) f_{\rm vol} \rho_{\rm m} R_{\rm m}\tilde{a}_{\rm m}^2}{L_\star(1-\alpha_{\rm m})}. \label{eq:tsubSimple}
\end{equation}
Equation \ref{eq:tsubSimple} assumes a constant orbital radius $\tilde{a}_{\rm m}$, as would be appropriate to a circular orbit.  If we instead integrate over an elliptical Keplerian orbit, the orbit-averaged sublimation time is more generally (e.g.~\citealt{Stone+15})
\begin{align}
\langle t_{\rm sub} \rangle =& \frac{2\pi (Q_{\rm vol}+ v_{\rm esc}^2/2) M_{\rm m} f_{\rm vol} \tilde{a}_{\rm m}^2 \sqrt{1-\tilde{e}_{\rm m}^2}}{L_\star R_{\rm m}^2 (1-\alpha_{\rm m})} \label{eq:tsubElliptic} \\
\approx & 1.4 \times 10^5~{\rm yr}~ \left( \frac{\tilde{q}_{\rm m}}{0.01\rm AU} \right)^{1/2} \left( \frac{\tilde{a}_{\rm m}}{\rm AU} \right)^{3/2} \left( \frac{M_{\rm m}}{10^{26}~{\rm g}} \right) \left( \frac{R_{\rm m}}{3\times 10^{8}~{\rm cm}} \right)^{-2}. \notag
\end{align}
In the second line we have assumed $\tilde{e} \approx 1$ and an exomoon with Callisto-like parameters, i.e. $M_{\rm m} \approx 1.1 \times 10^{26}~{\rm g}$, $R_{\rm m} \approx 2.4\times 10^8 ~{\rm cm}$, $f_{\rm vol} \approx 0.5$, and $\alpha_{\rm m} \approx 0.22$.  Note that for highly eccentric orbits, the majority of energy deposition from irradiation comes during brief pericenter passages.  As shown in Figure \ref{fig:histogram} and Table \ref{table:simresults}, the energy-limited sublimation timescale of the detached exomoons from our models span a large range, $\langle t_{\rm sub} \rangle \sim 10^{4}-10^{6.5}$ yr.

Tidal squeezing also contributes to heating the exomoon.  However, from \citet{Bodenheimer+01} (their Eqs.~2-3), the ratio of tidal heating to radiative heating is given by
\begin{equation}
    \frac{\langle\dot{E}_{\rm t}\rangle}{\langle\dot{E}_{\rm rad}\rangle} = \left(\frac{63\pi}{2Q}\right)\left(\frac{GM_\star^2\tilde{\Omega}_{\rm m}}{\tilde{a}_{\rm m}L_\star}\right)\left(\frac{\tilde{a}_{\rm m}}{R_{\rm m}}\right)^{-3}\left(\frac{\tilde{q}_{\rm m}}{\tilde{a}_{\rm m}}\right)^{1/2}\tilde{e}_{\rm m}^2\sqrt{1+\tilde{e}_{\rm m}},
\end{equation}
where $\tilde{\Omega}_{\rm m} \equiv (GM_{\star}/\tilde{a}_m^{3})^{1/2}$.  Taking $Q \sim 100$ for the exomoon, we find that $\langle\dot{E}_{\rm t}\rangle/\langle\dot{E}_{\rm rad}\rangle \sim 10^{-11}$ for characteristic parameters of the attached exomoons, rendering tidal heating negligible.

\subsection{Solid Particle Evolution and Stellar Dimming} 
\label{sec:solid}

Outgassing exomoon material will be comprised of solid particles with a range of sizes.  Consider first a single dust grain of size $b$ and bulk density $\rho_{\rm d} \approx 2$ g cm$^{-3}$ in an orbit about the star (of luminosity $L_{\star} \simeq 4.7L_{\odot}$, mass $M \simeq 1.43M_{\odot}$, and radius $R_{\star} \approx 1.58 R_{\odot}$) with a semi-major axis $a$, eccentricity $e$, and pericenter radius $q = a(1-e)$.

The outwards radiation force exceeds the inwards gravitational force above the Eddington luminosity, which for a spherical grain of cross section $\sigma = \pi b^{2}$ and mass $m_d = 4\pi b^{3}\rho_{\rm d}/3$ is given by
\be
L_{\rm edd} = \frac{4\pi G M_{\star}m_d c}{\sigma} =  \frac{16}{3}\pi b G M_{\star}\rho_{\rm d} c 
\ee
Equating $L_{\rm edd} = L_{\star}$ gives the maximum size for radiation blow-out,
\begin{align}
b_{\rm rad} =& \frac{3L_{\star}}{16\pi G M_{\star}\rho_{\rm d}c} \\ 
\approx& 0.95{\mu m} \left(\frac{L_\star}{4.7L_\odot} \right) \left(\frac{M_\star}{1.43M_\odot} \right)^{-1} \left(\frac{\rho_{\rm d}}{2~{\rm g~cm}^{-3}} \right)^{-1}. \notag
\end{align}
Small dust grains with $b \ll b_{\rm rad}$ may be present in the immediate vicinity of the outgassing exomoon, and could contribute to achromatic extinction of the star light during the dipping events \citep{Deeg+18,Schaefer+18}.  However, such small grains will not remain in orbit around the star and therefore cannot contribute to its long-term secular dimming.

By contrast, dust grains from the exomoon with $b \gg b_{\rm rad}$ are not strongly affected by radiation pressure and therefore will inherit initial orbits similar to that of the exomoon.  These larger solids will be dragged closer to the star over longer timescales by Poynting-Robertson (PR) drag \citep{Burns+79}.  In the limit of circular orbits, the timescale for PR drag is given by
\be
t_{\rm PR}^{\rm circ} = \frac{4 \rho_{\rm d}c^{2}b a^{2}}{3 L_{\star} Q_{\rm PR}} \approx 90\,\,{\rm yr} \,\, \left(\frac{b}{b_{\rm rad}}\right)\left(\frac{a}{\rm AU}\right)^{2}, \label{eq:tPRCirc}
\ee
where $Q_{\rm PR}$ is a dimensionless transmission coefficient hereafter taken to be unity.  However, in general the dust streams from exomoon outgassing will inherit the high eccentricity of their parent body (Fig.~\ref{fig:dust}), rendering Eq. \ref{eq:tPRCirc} inaccurate.  In the limit of a nearly radial orbit (i.e. $1-e \ll 1$), the PR inspiral time is substantially shorter, namely 
\be
t_{\rm PR}^{\rm rad} = \frac{16\sqrt{2} \rho_{\rm d}c^{2} b a^{1/2}q^{3/2}}{15 L_{\star}Q_{\rm PR}} = \frac{4\sqrt{2}}{5}\left(\frac{q}{a}\right)^{3/2} t_{\rm PR}^{\rm circ}. \label{eq:tPRe}
\ee
Dust will not survive long if its pericenter radius $q$ lies inside the sublimation radius,
\be
R_{\rm sub} = (T_{\star}/T_{\rm rock})^{2}R_{\star} \approx 16R_{\odot} \approx 0.074{\rm AU},
\label{eq:Rsub}
\ee
at which $T_{\rm eq} \gtrsim T_{\rm rock}$ (Eq.~\ref{eq:Teq}).  Using this minimum pericenter for liberated dust (i.e. requiring $q> R_{\rm sub}$), the ratio $t_{\rm PR}^{\rm rad} / t_{\rm PR}^{\rm circ}$ can in principle be as small as $\sim 10^{-3}$.  

We note that the relatively short PR drag timescale implies that injected dust will remain roughly coplanar: after the destruction of the parent planet, there are few sources of nodal precession in the system, and the dominant one is likely to be a (misaligned) spin-induced quadrupole moment on \tabby{}.  The nodal precession rate for dust from the detached exomoon (in the $q\ll a$ limit) is 
\begin{equation}
    \dot{\Omega}_{\rm rot} = \frac{-k_{\rm 2, \star}\Omega_\star^2 R_\star^5}{8(GM_\star a^3)^{1/2}q^2} \cos(i),
\end{equation}
where the angle $i$ is the misalignment between the stellar spin axis and the dust orbit's normal vector.  Generally, $|t_{\rm PR}^{\rm rad}\dot{\Omega}_{\rm rot}| \ll 1$, implying that dust inspirals into \tabby{} long before it is able to precess out of its birth plane.  A similar statement holds true for the advance of the dust pericenter, since $\dot{\Omega}_{\rm rot} \sim \dot{\omega}_{\rm rot}$, PR drag alone does not cause apsidal (or nodal) precession \citep{Veras+15}, and we have already seen that GR apsidal precession is subdominant to that from the rotational oblateness of \tabby{} (\S \ref{sec:Kozai}).  The relative unimportance of apsidal precession during the PR inflow period means that the dust orbits will not axisymmetrize, but instead will trace out the elliptical geometry they have inherited from the exomoon, with some natal spread in orbital elements imparted by the outgassing process.  One may determine the fractional spread in specific angular momentum for dust emitted at a radius $r$ with a velocity $\approx v_{\rm esc}$ as $\delta J/\tilde{J}_{\rm m} \approx r v_{\rm esc}/\sqrt{2GM_\star \tilde{q}_{\rm m}}$.  When $r \sim \tilde{q}_{\rm m}$, the ratio $\delta J/\tilde{J}_{\rm m} \ll 1$ and dust will be born into coaligned elliptical orbits, with a small spread in orbital pericenter and longitude of pericenter.  The subset of dust emitted on scales $r\gtrsim 1~{\rm AU}$ may satisfy $\delta J/\tilde{J}_{\rm m} \gtrsim 1$, and therefore be born into a more axisymmetric configuration, but as the bulk of energy-limited outgassing occurs near pericenter (Eq. \ref{eq:tsubSimple}), we expect this component of the dust disk to be small in mass.  The overall dust disk geometry is schematically illustrated in Fig. \ref{fig:cartoon}, and resembles that of Fig. 11 in \citet{Wyatt+18}, which may allow for greatly reduced infrared emission (we return to this point in \S \ref{sec:discussion}).

In order to evaluate the ability of liberated dust to explain secular evolution in the observed stellar luminosity, we have made a simplified geometric model for long-lived dust populations.  We begin by considering a large population of dust grains with constant $b$, $a$, and $e$, traveling on Kepler orbits.  For ease of calculation, we assume an axisymmetric dust disk with constant dimensionless aspect ratio $h \ll 1$.  We assume that long-lived dust populations become fully phase-mixed in true anomaly $f$; this will happen quickly because of the energy spread the dust grains are born with (as we show later in Eq.~\ref{eq:tmix}).  Mixing is not expected in $\omega$ due to the lack of apsidal precession described above.  Because the dust flows we expect are coaxial ellipses rather than axisymmetric annuli, the following toy model is inaccurate by geometric factors of order unity, but gives approximate estimates for the dynamics of a PR-mediated inflow.

Because the orbits are Keplerian, their number density profile $n_{\rm d}(R) \propto R^{-3/2}$ (and since $h \ll 1$, $n_{\rm d}$ depends weakly on vertical distance up to a disk scale height $H=hR$). The dust density profile can be normalized using the total number of grains in the disk, $N_{\rm d}$, so that
\begin{equation}
    n_{\rm d}(R) = \frac{3N_{\rm d}}{16\pi \sqrt{2} a^3} \left( \frac{R}{a} \right)^{-3/2}.
\end{equation}
Naively, the scale-height of the disk will be set by the escape speed of matter from the outgassing exomoon relative to its orbital velocity,
\be
h \approx \frac{v_{\rm esc}}{v_{\rm orb}} \approx 0.06\left(\frac{a}{\rm AU}\right)^{1/2},
\label{eq:hovera}
\ee
where we have used $v_{\rm esc} \approx 2$ km s$^{-1}$ as the value for Io (Table \ref{table:moons}), and
\be
v_{\rm orb} = \left(\frac{GM_{\star}}{a}\right)^{1/2}
\ee
as a characteristic orbital velocity of the exomoon near its semimajor axis $a$.  If the majority of the outgassing occurs during pericenter passage, then $h$ will be much smaller.  Note that
\be
\frac{hq}{R_{\star}} \approx 1.2 \left( \frac{q}{0.1{\rm AU}}  \right)\left(\frac{a}{\rm AU}\right)^{1/2} \left( \frac{v_{\rm esc}}{2~{\rm km~s}^{-1}} \right)
\label{eq:covering}
\ee
and so we expect that, even at the smallest realistic pericenters, the debris cloud will cover most if not all of the stellar disk if seen edge-on.

In order to block a fraction $f \lesssim 1$ of the star's light, then in the optically thin limit, the minimum optical depth of the dust grains along our line of sight at any time must be $\tau = f$, where $\tau = \pi b^2 S$, and the areal density of grains along our line of sight is $S = \int n(R){\rm d}R$.  Interestingly, for $n_{\rm d}(R) \propto R^{-3/2}$, dust absorption is dominated by the small number of grains at {\it pericenter}, not the much larger number at orbital apocenter.  The optically thin $\tau$ is thus
\begin{equation}
    \tau = \frac{3N_{\rm d}b^2}{8\sqrt{2} a^{3/2}}\left(q^{-1/2} - (2a)^{-1/2}\right).
\label{eq:tauperi}
\end{equation}

Putting everything together, and assuming $q \ll a$, the minimum mass in grains of size $b$ that is required to explain an obscured fraction $f = \tau$ is given by
\begin{align}
    M_{\rm min} &= \frac{32\pi \sqrt{2}}{3} \tau \rho_{\rm d}b a^2 q^{1/2} \label{eq:MMin}\\
    &\approx 2.4 \times 10^{-4} M_{\rm Io} \left( \frac{\tau}{0.1} \right) \left( \frac{b}{b_{\rm rad}} \right) \left( \frac{a}{\rm AU} \right)^{3/2} \left( \frac{q}{0.1{\rm AU}} \right)^{1/2} \notag
\end{align}
Eqs. \ref{eq:MMin} is a true minimum mass requirement because we have assumed that all dust grains have the same size, and have normalized this size to $b= b_{\rm rad}$, itself the smallest dust that can avoid radiation pressure blowout.

If a steady-state between sublimation and PR-mediated depletion is reached, the required minimum mass loss rate (assuming that $q\ll a$ and therefore Eq. \ref{eq:tPRe} applies) is
\begin{align}
\dot{M}_{\rm ss} =& \frac{M_{\rm min}}{t_{\rm PR}^{\rm rad}} = \frac{10}{9} \frac{a}{q}\frac{L_\star Q_{\rm PR}}{c^2}\tau \label{eq:MDotMin} \\
\approx& 7.9\times 10^{-6} M_{\rm Io}\,{\rm yr^{-1}}\left(\frac{\tau}{0.1}\right)\left(\frac{a/q}{10}\right).\notag
\end{align}
Remarkably, all absolute physical scales, aside from the luminosity of the star $L_\star$, drop out of Eq. \ref{eq:MDotMin}, which does not even depend on grain size $b$.  This will facilitate our generalization to a distribution of dust grain sizes, ${\rm d}N_{\rm d}/{\rm d}b$.  

In considering a grain size distribution, we need to distinguish between grains small enough to have reached a steady-state PR flow, and those big enough that they have not.  For an exomoon that has been evaporating for a time $t_{\rm age}$, the critical grain size bounding the steady-state population is 
\begin{eqnarray}
    b_{\rm ss} &=& \frac{15L_\star Q_{\rm PR} t_{\rm age}}{16\sqrt{2}\rho_{\rm d} c^2 a^{1/2} q^{3/2}}  \nonumber \\
&\approx& 0.03~{\rm cm} \left(\frac{t_{\rm age}}{10^3~{\rm yr}} \right) \left(\frac{a}{\rm AU} \right)^{-1/2} \left(\frac{q}{0.1 \rm AU} \right)^{-3/2}. \label{eq:bss}
\end{eqnarray}
Dust grain sizes are often distributed as power laws in nature, with ${\rm d}N_{\rm d}/{\rm d}b \propto b^{-\delta}$ with $b_{\rm min}<b<b_{\rm max}$.  If $\delta<4$, the total dust mass $M_{\rm d}$ is dominated by grains with $b\sim b_{\rm max}$; if $\delta>3$, the total dust area $A_{\rm d}$ is dominated by grains with $b\sim b_{\rm min}$.  Observed dust populations produced by cometary evaporation or collisional cascades typically have $\delta \approx 3.5$ \citep{Dohnanyi69}.  

If we define an area-to-mass ratio $\Upsilon \equiv A_{\rm d}/M_{\rm d}$, we can derive simple corrections to the above formulae.  The area-to-mass ratio of a dust population is maximized for a homogeneous population of dust grains with size $b = b_{\rm rad}$, which has $\Upsilon_{\rm max} = 3 / (\rho_{\rm d}b_{\rm rad})$.  We can therefore compute corrections to $M_{\rm min}$ in the following way: the true dust mass required to explain an obscured fraction $f$ is 
\be M_{\rm true} = M_{\rm min}(b_{\rm rad}) \times \frac{3}{\rho_{\rm d} b_{\rm rad} \Upsilon}.  \nonumber \ee
If dust is injected with a continuous power-law distribution, then
\begin{equation}
    \Upsilon = \frac{3}{\rho_{\rm d}}\frac{4-\delta}{3-\delta} \frac{b_{\rm max}^{3-\delta} - b_{\rm min}^{3-\delta}}{b_{\rm max}^{4-\delta} - b_{\rm min}^{4-\delta}} \approx \frac{3}{\rho b_{\rm max}} \frac{4-\delta}{\delta -3} \left(\frac{b_{\rm min}}{b_{\rm max}} \right)^{3-\delta},
\end{equation}
where the approximate equality assumes $3<\delta<4$.  While this simple power law is a plausible estimate for the {\it injection spectrum} of dust from a sublimating body, over longer timescales it will be sculpted by PR drag.  

Let us assume that dust particles are being injected at the differential rate ${\rm d}\dot{N}_{\rm d} / {\rm d}b \propto b^{-\delta}$.  Dust grains with size $b>b_{\rm ss}$ have not yet reached a PR steady state, and so their differential size distribution will match the differential injection spectrum.  However, smaller dust grains will have come into steady state, and so their ${\rm d}N_{\rm d} / {\rm d}b \sim {\rm d}\dot{N}_{\rm d} / {\rm d}b \times t_{\rm PR}^{\rm rad}(b) \propto b^{1-\delta}$.  The two power laws smoothly meet at $b= b_{\rm ss}$, the critical grain size {\it which dominates the area budget}, so long as $\delta < 4$.

\section{Discussion and Application to \tabby}
\label{sec:discussion}

Exoplanet-star collisions may produce prompt electromagnetic transients (e.g.~\citealt{Bear+11,Metzger+12}), and the thermalization of orbital energy into the star's outer layers will temporarily increase the star's luminosity above its main sequence level \citep{Metzger+17} for a period of years (for terrestrial exoplanets) to millenia (for Jovian exoplanets).  The dynamical models we have developed so far suggest that these events can also be accompanied by longer-lived signatures related to any exomoons tidally detached from the consumed planet.  In this section, we discuss both transits from debris clouds generated by violent evaporation of exomoon surface layers and secular variations in the luminosity of the host star; in both cases, we focus particularly on \tabby{}.

\subsection{Debris Clouds and their Transits}
\label{sec:clouds}
Exoplanets receiving sufficiently strong irradiation from their host star will lose mass through evaporation.  For gas giants, this mass loss will take the form of thermal (Jeans) escape from the upper atmosphere \citep{VidalMadjar+03}.  A similar process occurs for solid exoplanets, although here the radiative heating must first sublimate their surface layers.  Several observed stars exhibit short-period, asymmetric, variable-depth transits \citep{Rappaport+12, Rappaport+14, SanchisOjeda+15}.  The origin of these transits has been interpreted as a dusty debris cloud escaping from an evaporating solid world.  However, these systems are not precise analogs to the scenario outlined in this paper, as they involve (i) exoplanets on quasi-circular (rather than nearly radial) orbits, and (ii) direct sublimation of volatile-poor, rocky worlds \citep{vanLieshout+14}. 

When fractional sublimation rates are relatively low, and surface escape speeds are relatively high, one may model mass loss with an atmospheric escape formalism \citep{PerezBecker&Chiang13}, but this formalism breaks down in the opposite (``catastrophic mass loss'') regime that applies to the smaller exomoons we are considering.  Instead, we have estimated the {\it minimum}, energy-limited time for a solid body on an eccentric orbit to fully evaporate in Eq. \ref{eq:tsubElliptic}.  The true evaporation time is likely longer due to the effects of radiative cooling and self-shielding, which we have negleted.

Following several previous works on \tabby~(e.g.~\citealt{Boyajian+16,Bodman&Quillen16,Neslusan&Budaj17}), we hypothesize that the observed dip structures are produced in the stellar light curve as an outgassing solid body (in our model, a tidally detached exomoon) brings it across the observer's line of site.  However, the precise number of distinct exomoons which are required to explain the data is unclear; for instance, \citet{Neslusan&Budaj17} argue that the oberved phenomenology can be explained by four dust-enshrouded bodies.  %Also note that, although no precise periodicities are observed in the dipping behavior, this does not necessarily constrain a periodic orbit for the underlying body.  

The detailed structures of the observed light curve dips will depend on the geometry of the dusty debris flows leaving the detached exomoon.  While unbound dust is produced by outgassing and sublimation on the surface of the exomoon, dust orbits are subsequently shaped by the competition between stellar gravity and various perturbing forces, such as radiation effects and the gravitational field of the exomoon itself.  However, considering only the gravitational force of the central star (which is expected to dominate for grains much larger than the radiation blow-out limit), we can make a simple estimate of the timescale for material released from the exomoon to phase-mix along the orbit and thus lose its spatially-coherent structure surrounding the exomoon.  

As discussed earlier, debris will leave the exomoon with a velocity dispersion approximately equal to its escape velocity $v_{\rm esc} = (2GM_{\rm m}/R_{\rm m})^{1/2}$ and a specific energy spread $\delta\epsilon \sim v_{\rm esc}v_{\rm orb}(r)$, for material released at a radius $r$ from the central star (typically, radii $r \sim \tilde{q}_{\rm m}$ will source most of the mass loss).  
There is consequently a spread in orbital periods $\delta P / \tilde{P}_{\rm m} = (3/2)(\delta \epsilon / \tilde{\epsilon}_{\rm m})$, where $\tilde{P}_{\rm m}$ and $\tilde{\epsilon}_{\rm m}$ are the post-detachment period and orbital energy of the exomoon, respectively.  The number of orbits required to dynamically phase mix a ``clump'' of evaporated gas and dust into a smooth, elliptical ring is therefore
\begin{align}
    N_{\rm mix} =& \frac{1}{3\sqrt{2}}\left( \frac{M_\star}{M_{\rm m}} \right)^{1/2} \left( \frac{r R_{\rm m}}{\tilde{a}_{\rm m}^2} \right)^{1/2} \label{eq:tmix} \\
    \approx& 1.0 \left( \frac{M_{\rm m}}{5\times 10^{25}~{\rm g}} \right)^{-1/2} \left( \frac{R_{\rm m}}{2 \times 10^8~{\rm cm}} \right)^{1/2} \left( \frac{r}{0.1~{\rm AU}} \right)^{1/2} \left( \frac{\tilde{a}_{\rm m}}{2~{\rm AU}} \right)^{-1}.    \notag
\end{align}
In the second line, we have assumed the stellar properties of \tabby{}.  The speed of phase mixing, and therefore the clumpiness of the outgassing debris, depends on both the size of the exomoon and its post-detachment orbital properties.  Smaller and more tightly bound exomoons will produce debris with a smaller fractional energy spread, i.e. the ``clumpier,'', $N_{\rm mix} > 1$ regime.  Conversely, larger and more loosely bound exomoons can produce debris with a large fractional energy spread ($\delta \epsilon / \tilde{\epsilon}_{\rm m} > 1$), in which case the ejected dust and gas sprays out onto orbits with a wide range of eccentricities and semimajor axes.  Clumpiness also depends on the radius $r$ along the orbit at which the debris in question is emitted.

In the $N_{\rm mix} > 1$ regime, material produced during the $(i)$th pericenter passage will retain some coherence on the $(i+1)$th pericenter passage, and possibly subsequent ones as well.  This regime should produce transits with rich ingress and egress structure: the deepest portion of the transit is likely to involve freshly-sublimated material in close proximity to the exomoon itself (both because that material has not had time to shear out, and because small grains with $b < b_{\rm rad}$ have not had time to blow out of the system), but there may be greatly extended, intermediate-timescale structure on the ingress and egress, produced by dusty debris tails from prior pericenter passages that have not yet fully phase mixed.  Conversely, the $N_{\rm mix} < 1$ regime is likely to produce a simpler transit, with most of the obscuring material just evaporated from the detached exomoon.

In principle, volatile-rich exomoons may begin to produce dust clouds from massive sublimation {\it prior} to tidal detachment from their parent exoplanet.  As we illustrate in Fig. \ref{fig:collision}, detachment often occurs once the exoplanetary pericenter $q_{\rm p} \lesssim 10 R_\star$, and therefore, many of the pre-detachment pericenter passages will be accompanied by sizable sublimation and mass loss of still-bound exomoons.  However, the dust and gas produced from pre-detachment orbits will not be able to freely expand, and will likely remain gravitationally bound to the parent exoplanet (typically, $v_{\rm th}^2 \ll GM_{\rm p} / a_{\rm m}$, at least for exomoon systems similar to those around Solar gas giants).  While substantial dips in the stellar light curve are possible for dust clouds in ``planetocentric'' orbits, the pre-detachment phase of a KL cycle does not seem to naturally explain the secular dimming of \tabby{} via long-term debris evolution, as we explore in the next subsection.  Sublimation of pre-detached exomoons might, however, explain unusual ``dipper'' stars showing deep, irregular transits that lack the secular dimming seen in \tabby{}; see, for example, \citet{Rappaport+18, Ansdell+19, Schaefer19}.

\subsection{Secular Dimming}
The lifetime of the disk of solid particles created by exomoon outgassing is given by the longer of the moon sublimation time (Eq.~\ref{eq:tsubElliptic}) and the timescale to drain under the influence of Poynting-Robertson drag.  To provide a lower limit on the latter, we assume that the maximum injected particle size, $b_{\rm max}$ is below the critical size $b_{\rm ss}$ (Eq.~\ref{eq:bss}) needed to establish the steady accretion rate $\dot{M}_{\rm ss}$ (Eq.~\ref{eq:MDotMin}).  This gives a minimum accretion time through the solid disk of
\be
t_{\rm solid} \approx \frac{M_{\rm m}}{\dot{M}_{\rm ss}} \approx 1.2\times 10^{5}\,{\rm yr}\,\left(\frac{M_{\rm m}}{M_{\rm Io}}\right)\left(\frac{\tau}{0.1}\right)^{-1}\left(\frac{a/q}{10}\right)^{-1}, \label{eq:tsolid}
\ee
where the radial optical depth through the disk is normalized to the value $\tau \gtrsim 0.1$ required to explain the $\gtrsim 10\%$ secular dimming of \tabby~ (\citealt{Schaefer16,Montet&Simon16,Simon+18,Schaefer+18}).  This equation was derived under the assumptions of steady-state optically-thin PR flow.  The steady-state assumption is valid provided that $b_{\rm max}$ does not exceed the critical size (Eq.~\ref{eq:bss})
\be
b_{\rm ss}(t_{\rm age} = t_{\rm solid}) \approx 0.3~{\rm cm}\,\left(\frac{M_{\rm m}}{M_{\rm io}}\right)\left(\frac{\tau}{0.1}\right)^{-1}\left(\frac{a}{\rm AU} \right)^{-3/2} \left(\frac{q}{0.1 \rm AU} \right)^{-1/2}
\ee
It is not immediately obvious what the large end of the dust grain size distribution from a massively outgassing exomoon crust will be.  We can crudely calibrate our expectations by considering the dust grain distributions found in rapidly sublimating Solar System comets.  For example, the Deep Impact space mission to the comet Temple 1  (\citealt{AHearn+05,Gicquel+12}) found maximum particle sizes of $b_{\rm max} < 100\mu$m, which safely satisfy this limit.  The PR drag timescale we have assumed (Eq.~\ref{eq:tPRe}) will also be modified if the solid disk is optically thick (e.g.~\citealt{Rafikov11}); however, as optical depth effects will only act to suppress the drag rate, $t_{\rm solid}$ is a conservative minimum timescale for the disk draining. 

Figure \ref{fig:histogram} shows histograms of $\langle t_{\rm sub} \rangle$ (Eq.~\ref{eq:tsubElliptic}) and $t_{\rm solid}$ (Eq.~\ref{eq:tsolid}) for the population of detached exomoon orbits from our Fiducial and Small exoplanet models, assuming an exomoon mass of $M_{\rm m} = 10^{26}$ g.  Although we find minimum sublimation times in the range $\sim 10^{4}-10^{6.5}$ yr, most are in the range $\gtrsim 10^{5.5}$ yr when we restrict ourselves to just the subset of exomoons which obey observational constraints on the orbits of the outgassing bodies responsible for the dimming of~\tabby~ \citep{Schaefer+18}.  We find a smaller range for the solid accretion timescale, $t_{\rm solid} \sim 10^{4}-10^{5}$ yr.  The fact that $t_{\rm solid} \lesssim \langle t_{\rm sub} \rangle$ shows that the solid disk lifetime is controlled by the rate at which the exomoon evaporates, rather than the timescale for processing liberated material through the solid disk.\footnote{If the opposite were instead true (namely, $t_{\rm solid} \gg \langle t_{\rm sub} \rangle$) then we would expect a large population of stars with opaque debris disks from long-ago evaporated exomoons without ongoing dipping behavior.}  

Equating the sublimation time $\langle t_{\rm sub} \rangle$ to the solid draining time $t_{\rm solid}$ gives an estimate of the optical depth through the solid disk in sublimation/accretion equilibrium,
\begin{eqnarray}
\tau_{\rm eq} &=& \frac{9}{20\pi\sqrt{2}}\frac{(1-\alpha_{\rm m})}{Q_{\rm PR}f_{\rm vol}}\left(\frac{c^{2}}{Q_{\rm vol}}\right)\left(\frac{q^{1/2}R_{\rm m}^{2}}{a^{5/2}}\right) \nonumber \\
&\approx& 0.54\left(\frac{q/a}{0.1}\right)^{1/2}\left(\frac{R_{\rm m}}{3\times 10^{8}{\rm cm}}\right)\left(\frac{a}{\rm 3\,AU}\right)^{-2}, \label{eq:taueq}
\end{eqnarray}
where in the final line we have again adopted fiducial values $Q_{\rm PR} = 1$, $f_{\rm vol} =0.5$, $\alpha_{\rm m} = 0.22$, $Q_{\rm vol} \approx 3\times 10^{10}$ erg g$^{-1}$.  Thus, for typical values $q \sim a/10$ and $a \lesssim $ 3 AU, we expect a moderately opaque disk $\tau = \tau_{\rm eq} \gtrsim 1$, which combined with the expected scale-height of the solid disk (Eq.~\ref{eq:covering}) is consistent with the order-unity fractional secular dimming of \tabby~\citep{Schaefer16}.

Given these estimates for the lifetimes of detached exomoons and their obscuring material, we can now estimate the fraction of stars that must undergo exoplanet-star collisions in order to explain the occurrence of an object like \tabby~in the Kepler field (\citealt{Lacki16}).  Our numerical results show that a fraction $f_{\rm sur} \sim 0.1$ of exomoons will survive the destruction of their host exoplanet (Table \ref{table:simresults}).  Typically, most of these stable exomoon orbits have sufficiently small pericenters to make massive outgassing and production of dusty debris inevitable, provided the exomoon has volatile-rich surface layers.  If each exoplanet hosts $N_{\rm m}$ major exomoons, then we expect that a fraction $N_{\rm m}f_{\rm sur}$ of exoplanet-star encounters will result in at least one exomoon orbiting the star.  The typical exomoon or solid disk lifetime set by evaporation, $t_{\rm life} \sim 10^{6}$ yr (Fig.~\ref{fig:histogram}), corresponds to a fraction $f_{\rm life} = t_{\rm sub}/t_{\rm MS} \sim 10^{-3}$ of the total main-sequence lifetime $t_{\rm MS} \approx 2$ Gyr of an F-star.  Given the vertical aspect ratio of the solid disk $h \sim 0.1$ (see Eq.~\ref{eq:hovera} and surrounding discussion), we expect that a fraction $f_{\rm h} \sim 0.1$ of observers will possess the correct orientation in the plane of the exomoon's orbit to observe both dipping and secular dimming.  Finally, there are approximately $N_{\rm tot} \sim 10^{5}$ stars total in the {\it Kepler} sample (though only $N_{\rm F} \approx 5000$ stars have effective temperatures similar to \tabby).  Also factoring into consideration is the precise number of surviving exomoons required to explain the dipping behavior of \tabby~ (\S\ref{sec:clouds}).

Given the observation of a single star similar to \tabby, we therefore expect that a fraction
\be
f_{\rm coll} \sim N_{\rm m}f_{\rm sur}f_{\rm life}f_{\rm h}\frac{N_{\rm tot}}{1\,{\rm \tabby}} \sim 1\times N_{\rm m}\left(\frac{f_{\rm sur}}{0.1}\right)\left(\frac{f_{\rm h}}{0.1}\right)  \label{eq:Ncoll}
\ee
of stars must over their lifetimes undergo an exoplanet destruction or close encounter events.  At face value the detached exomoon hypothesis therefore provides a reasonable explanation for \tabby~if an order unity fraction of stars experience a strong encounter with a exomoon-bearing exoplanet at some point in its main-sequence lifetime.  Is this reasonable?  High-eccentricity migration has long been viewed as a plausible primary formation channel of Hot Jupiters (e.g.~\citealt{Rasio&Ford96}), and the specific mechanism we have investigated here (KL cycles) certainly occurs around some fraction of star-exoplanet systems with outer companions \citep{Wu+07}.  While the fraction of Solar type stars hosting Hot Jupiters is only $\sim 1\%$ \citep{Wright+12}, the exoplanet fraction as a whole is higher around more massive stars \citep{Johnson+10}.  However, it is unclear what fraction of giant stars placed on close encounters with their host stars are able to successfully circularize into stable orbits, versus being tidally disrupted or directly colliding with the star.  \citet{Stephan+18} argue that about 25\% of gas giants orbiting A-type stars will be destroyed by their stars during the main sequence lifetime as a result of KL-induced encounters, an estimate not far from that required from equation (\ref{eq:Ncoll}).

\section{Conclusions}
\label{sec:conclusions}

Icy moon systems are ubiquitous among the giant planets in our Solar System, the prototypical example being the Galilean moons of Jupiter.  If the eccentricity of an analogous exoplanet is increased to a sufficiently high value by gravitational interactions with an outer perturber (Fig.~\ref{fig:KLtime}), then its exomoons will be removed from the exoplanet's grip by stellar tidal forces and placed into orbit around the central star.  In some cases this will take place during high-eccentricity excursions of the exoplanet's orbit that do not immediately result in its destruction (Fig.~\ref{fig:collision}), but in many other cases, exomoon detachment more immediately precedes the exoplanet's demise.  

Our few-body simulations demonstrate that a sizable fraction ($\sim 10\%$) of the detached exomoons can survive further gravitational interaction with the exoplanet and end up on stable eccentric orbits (Table~\ref{table:simresults}).  Their average eccentricity and semi-major axes are predicted reasonably well by analytic considerations ($\S\ref{sec:analytic}$); however, substantial dispersion in these properties is produced by ongoing perturbations from the exoplanet between the time of detachment and the time of the exoplanet's destruction (Fig.~\ref{fig:orbitproperties}).  While we have not conducted an exhaustive parameter study, which would entail consideration of a great number of possible exoplanetary architectures, our results are a proof-of-concept demonstration of how many exomoons will survive the destruction of their host gas giants.  We have also demonstrated that a collision of the exomoon with its host exoplanet due to secondary KL-induced oscillations from the central star are unlikely to destroy the moons prior to tidal detachment (as a result of detuning due to the spin-induced quadrupole moment of the planet; Fig.~\ref{fig:secondaryKL}). 

After detachment, volatile-rich exomoons find themselves exposed to much stronger stellar irradiation than they experienced in their birth environment.  The exomoons' highly eccentric orbits typically reside outside the radius of solid sublimation, but entirely inside the ice line (Fig.~\ref{fig:dust}).  In a somewhat extreme analogy with comet pericenter passages in our own Solar System, evaporation of ices on the exomoon surface will likely result in out-gassing of small particulates that shroud the exomoon in a dusty debris cloud (Fig.~\ref{fig:cartoon}).  

As the exomoon passes in front of the observer's line of sight, extended, opaque clouds of dusty debris could generate deep localized minima in the stellar light-curve.  Weaker dips may be generated by extended debris clouds, generated by outgassing during prior orbits, that have not yet had time to phase mix into a more coherent debris disk.  While small grains (those less than one micron in size) are quickly removed by radiation pressure, larger grains are less affected and will initially settle into eccentric orbits inherited from the parent exomoon.  These larger particles will then phase mix and accrete onto the star over much longer timescales, via PR drag ($\S\ref{sec:solid}$).  Provided that most of the solid disk's mass resides in relatively small grains less than a centimeter in size, the steady-state PR drag accretion rate (Eq.~\ref{eq:MDotMin}), and thus its draining time (Eq.~\ref{eq:tsolid}), are remarkably insensitive to variables other than the total exomoon mass.  If solid particle creation through exomoon sublimation balances PR accretion, then the optical depth of stellar light passing radially through the solid disk is of order unity for characteristic parameters (Eq.~\ref{eq:taueq}).

\citet{Simon+18} and \citet{Schaefer+18} show that the secular ``dimming" of \tabby~is not monotonic, but instead the light curve occasionally shows brief flux increases.  In our picture, optical depth variability can be naturally produced by variability in the ``injection rate'' of mass from the evaporating exomoon, e.g. from compositional gradients or discontinuities in the evaporating surface layers.  Any mismatch between the rate of dust injection and the steady-state Poynting-Robertson flow rate through the disk will produce variability in the total optical depth of the solid disk.  Since the optical depth through the solid disk is dominated by those particles on orbits with the shortest pericenter distances (corresponding to the radius of solid sublimation, $R_{\rm rock} \lesssim 0.1$ AU; see Eq.~\ref{eq:tauperi} and surrounding discussion), then order-unity variations in the amount of secular-timescale obscuration could take place on timescales as short as (Eq.~\ref{eq:tPRe})
\be
t_{\rm PR}^{\rm rad}(q \sim R_{\rm rock}) \sim 0.7\,\,{\rm yr}\,\,\left(\frac{b}{b_{\rm rad}}\right)\left(\frac{a}{1\,{\rm AU}}\right)^{1/2}, 
\ee
where $b$ is the grain size which dominates dust opacity normalized to the mimium value $b_{\rm rad}$ allowed from radiation blow-out.

Stellar light absorbed by the solid disk will be re-emitted at longer infrared wavelengths.  Again, since the optical depth through the solid disk is dominated by small pericenter radii close to the sublimation radius (Eq.~\ref{eq:tauperi}), most of the re-processed emission will be in the near-infrared (NIR) at wavelengths $\lambda \approx hc/(3kT_{\rm rock}) \approx 3\mu$m.  Although stringent limits exist on the presence of persistent NIR excess from \tabby~\citep{Lisse+15,Marengo+15}, such constraints do not necessarily rule out this picture as long as the optical depth of the re-processed infrared emission through the disk is also high (as is necessarily the case because the large grains of the solid disk $\gg b_{\rm min} \sim 1\mu$m place the absorption of even NIR emission squarely in the geometric optics limit), and provided the geometry is such that the NIR emission is redirected away from the Earth (see \citealt{Wyatt+18} for details).  Our general expectation, set by the lack of nodal and apsidal precession in the solid disk (see \S \ref{sec:solid}) is that the PR-driven dust inflow will be (i) coplanar and (ii) a set of nested, coaxial ellipses, rather than an axisymmetric structure.  This geometry is indeed similar to that invoked in \citet{Wyatt+18} to limit NIR emission.

Previous work by \citet{Metzger+17} attributed the secular dimming behavior of \tabby~to changes in the intrinsic luminosity of the star as it returned to steady-state following the sudden injection of energy by the exoplanetary collision.  However, the relatively short timescale $\sim 10-1000$ yr of the predicted fading, demanded an uncomfortably high rate of exoplanet-star collisions.  The scenario presented here alleviates this tension because the solid disk is present around the star for the evaporation lifetime of the exomoon, which can be $\sim 100-1000$ longer for exomoon masses comparable to the major moons in our Solar System (Fig.~\ref{fig:histogram}).  If close encounters with moon-bearing exoplanets are a relatively common occurrence in the main sequence lifetimes of F-type stars, then the discovery of \tabby~would not be a large statistical anomaly (Eq.~\ref{eq:Ncoll}).

Although the main application of our study is to \tabby, the general scenario outlined here may be relevant to other stellar systems demonstrating unusual dipping lightcurve behavior, such as the ``dipper" stars showing deep, irregular transits which lack the secular dimming behavior seen in \tabby~\citep{Rappaport+18,Ansdell+19,Schaefer19}.  Photoevaporation of volatile material from the exomoon system {\it prior to tidal detachment from the exoplanet}, could generate a disk of dusty material which would remains gravitationally bound in orbit about the exoplanet.  Such a disk could in principle block light from the star as the eccentric orbit of exoplanet crosses the disk of the star, generating dips in the stellar light curve but without placing solid debris into a permanent orbit about the star. 

Our work has made a number of assumptions that should be relaxed in future work.  For instance, the true evaporation time of the exomoons will likely exceed our estimate in Eq. \ref{eq:tsubElliptic}, as we have neglected at least two potentially important effects: dust-gas coupling and self-shielding.  However, a longer exomoon survival timescale would just strengthen our arguments, by permitting a higher probability of {\it Kepler} catching such a system in its current state.  

One may also speculate that similar exomoon detachment events could take place during post-main sequence evolution, once the central star has become a white dwarf.  As stars evolve off the main sequence and lose significant fractions of their mass, associated exoplanetary systems can undergo massive rearrangements due to expanding the expanding widths of Hill spheres and mean motion resonances.  In many cases, exoplanets will be ejected or destroyed as the result of strong scatterings \citep{Veras&Tout12, Veras+13}.  In other cases, however, an exomoon-rich gas giant on a wide orbit that did not previously experience strong KL cycles may become vulnerable to them: as one example of this, secular torques from {\it inner} planets may prevent outer ones from exciting each other to very high eccentricities, but this stabilizing effect disappears once the inner planets are engulfed by a giant-branch star \citep{Petrovich&Munoz17}.  We note, however, that typically low luminosities of white dwarfs pulls in their ice lines relative to the expected orbits of detached exomoons, reducing the evaporation rate from our scenario, relative to the main sequence case.  Future work will address the implications of orphaned exomoons for white dwarf metal pollution (\citealt{Farihi16}) and the observability of exomoons via white dwarf transits \citep{Cortes&Kipping18}.    

\section*{Acknowledgements}
We thank Tabetha Boyajian, Dan Clemens, Diego Mu\~noz, Brad Schaefer, Josh Simon, and Jason Wright for thoughtful comments on earlier versions of the manuscript.  This work was supported in part by the NASA Astrophysics Theory Program (grant number NNX17AK43G).

\bibliographystyle{yahapj}
\bibliography{main}

\end{document}